\documentclass[11pt]{article}
\usepackage
[pdftex,hyperindex=true,hyperfigures=true,pagebackref=true,bookmarks=true,pdftitle=
{TITLE},pdfsubject=
{},pdfauthor
=
{AUTHOR
},pdfkeywords={KEYWORDS},pdfpagemode=UseOutlines,menubordercolor=1 1 1,colorlinks=true,urlcolor=blue,citecolor=red,linkcolor=blue]{hyperref}

\usepackage{amsmath,amsfonts,amssymb,epsfig}
\usepackage{graphics}
\usepackage{graphicx}
\usepackage{array}
\usepackage{bbm}
\usepackage[T1]{fontenc}

\newcommand{\al}{\alpha}
\newcommand{\BE}{\begin{equation}}
\newcommand{\EE}{\end{equation}}
\newcommand{\la}{\lambda}
\newcommand{\ome}{\omega}

\newcommand{\EAL}{E^\al_f}
 
\newcommand{\RR}{{\mathbb R}}
\newcommand{\R}{{\mathbb R}}

\newcommand{\ZZ}{{\mathbb Z}}
\newcommand{\NN}{{\mathbb N}}

\newcommand{\pL}{\ell^{(p)}}
\newcommand{\EEE}{{\mathbb E}}

\newcommand{\ep}{\varepsilon}
\newcommand{\be}{\beta}

\newcommand{\La}{\Lambda}

\newcommand{\TMFD}{{T}_{mfd}}

\newcommand{\Hmin}{{h^{min}_f}}

\newcommand{\pfx}{p_f(x_0)}
\newcommand{\qfx}{q_f(x_0)}

\newcommand{\cu}{c_1 (p)}
\newtheorem{lem}{Lemma}
\newtheorem{coro}{Corollary}
\newtheorem{Theo}{Theorem}
\newtheorem{prop}{Proposition}
\newtheorem{defi}{Definition}
\newcommand{\BP}{\begin{prop}}
\newcommand{\EP}{\end{prop}}
\newcommand{\BC}{\begin{coro}}
\newcommand{\EC}{\end{coro}}
\newcommand{\BL}{\begin{lem}}
\newcommand{\EL}{\end{lem}}
\newcommand{\BD}{\begin{defi}}
\newcommand{\ED}{\end{defi}}
\newcommand{\BT}{\begin{Theo}}
\newcommand{\ET}{\end{Theo}}

\def\E{{\hbox{I\kern-.2em\hbox{E}}}}

\graphicspath{
{Figures/}
}

\newsavebox{\fmbox}
\newenvironment{fmpage}[1]
 {\begin{lrbox}{\fmbox}\begin{minipage}{#1}}
 {\end{minipage}\end{lrbox}\fbox{\usebox{\fmbox}}}

\author{     St\'ephane  Jaffard\thanks{ 
       Address: Laboratoire d'Analyse et de Math\'ematiques Appliqu\'ees, CNRS, UMR 8050, UPEC,  Cr\'eteil, France 
jaffard@u-pec.fr }, 
  Guillaume Saës\thanks{   Address: Laboratoire d'Analyse et de Math\'ematiques Appliqu\'ees, CNRS, UMR 8050, UPEC,  Cr\'eteil, France 
 and D\'epartement
        de Math\'ematique, Universit\'e de Mons, Place du Parc 20, 7000 Mons (Belgium) guillaume.saes@u-pec.fr},  Wejdene Ben Nasr\thanks{   Address: Laboratoire d'Analyse et de Math\'ematiques Appliqu\'ees, CNRS, UMR 8050, UPEC,  Cr\'eteil, France wejdene.nasr@u-pec.fr
 },\\ Florent Palacin \thanks{   Laboratoire de neurophysiologie et de biom\'ecanique du mouvement, 
        Institut des neurosciences de l'Universit\'e Libre de Bruxelles, Belgium  palacinflorent@gmail.com},  
         V\'eronique  Billat\thanks{ Universit\'e Paris-Saclay, Univ Evry, F-91000 Evry-Courcouronnes, France. veronique.billat@billatraining.com
}
}

 \date{} 
\title{  A review of univariate and multivariate  multifractal analysis  illustrated by  the analysis of marathon runners physiological data
 }

\begin{document}
\maketitle
{ \bf Abstract:}  We  review  the central results concerning  wavelet methods in  multifractal analysis, which consists in  analysis of the pointwise singularities of a  signal, and we describe its recent extension to  multivariate  multifractal analysis, which deals with the joint analysis of several signals; we focus on the mathematical questions that this new techniques motivate. We illustrate these methods by an application to data recorded on marathon runners.    
\vspace{0.5cm}


\clearpage
\newpage
\tableofcontents

\clearpage
\newpage

\section{Introduction } 

\label{intro}

Everywhere irregular signals are ubiquitous in nature: Classical examples are supplied by natural phenomena (hydrodynamic turbulence \cite{H18}, geophysics, natural textures  \cite{Johnson2014}), physiological data (medical imaging \cite{Arneodo2003a}, heartbeat intervals \cite{ABRY:2010:A}, E.E.G \cite{EMBC19_Brain}); they are also present in  human activity and technology (finance \cite{Bacry2010b}, internet traffic \cite{MandMemor}, repartition of population \cite{frankhauser1998fractal,SemecurbeHuman} , text analysis \cite{Leonarduzzi2017ICASSP}, art  \cite{Wendt2013a}).  The analysis of such phenomena requires the modelling by everywhere irregular functions, and it is therefore natural to use mathematical regularity parameters in order to classify such data, and   to study mathematical models which would fit their behavior.  Constructing and understanding the properties of such functions has been a major challenge in mathematical analysis for a long time:  Shortly after Cauchy gave the proper definition of a continuous function, the question of determining if a continuous function is necessarily differentiable at some points was  a major issue for a large part of the 19th century; though a first counterexample was found by Bolzano, his construction remained unknown from the mathematical community, and it was only in 1872,  with the famous Weierstrass functions  
 \BE \label{weier1}  { \cal W}_{a, \ome} (x) = \sum_{n =0}^{+ \infty} \frac{\sin ( a^n x)}{a^{\ome n}}  \hspace{15mm} \mbox{for} \hspace{6mm}  a >1   \hspace{6mm} \mbox{and } \hspace{6mm}  \ome \in (0,1), \EE
 that the problem was settled.  However, such constructions were considered  as weird  counterexamples, and  not representative of what is commonly met, both in mathematics and in applications.  In 
 1893, Charles Hermite  wrote to 
  Thomas Stieltjes: { \em I turn my back with fright and horror  to this lamentable plague: continuous functions without derivative}. The first statement that smooth or piecewise smooth functions were not adequate for modelling  natural phenomena but were rather exceptional came from physicists, see e.g. the introduction of the famous book of Jean Perrin ``Les atomes'', published in 1913.   On the mathematical side,  the  evolution  was slow: 
In 1931,    Mazurkiewicz   and Banach  showed that most continuous functions are nowhere differentiable (``most'' meaning here  that such functions form a residual set in the sense of Baire categories).  This spectacular result changed the perspective: Functions which  were considered as  exceptional and  rather pathological actually were the common rule, and smooth functions turn out to be exceptional.

A first purpose of multifractal analysis is to supply mathematical notions which allow to quantify  the irregularity of functions, and therefore yield quantitative tools  that can be applied to real life data in order to determine if they fit a given model, and, if it is the case, to determine the correct parameters of the model.   One can also be more ambitious and wonder which ``types'' of singularities  are present in the data, which may yield an important information of the nature of the signal; a typical example is supplied by { \em chirps} which are  singularities which behave like
\BE \label{chirp} 
 g (x)   =  | x-x_0|^\al \cos \left(\displaystyle\frac{1}{| x-x_0|^\be }\right) ,  \EE  
displaying fast oscillations near the singularity at $x_0$. Such singularities  are e.g. predicted by  some models of turbulence and therefore determining if they can be found in the recorded data in wind tunnels is an important issue in the understanding of the physical nature of turbulence. 

A first step in this program was performed by A. Kolmogorov in  1941  \cite{Kol41}.  
 Let $f: \RR^d \rightarrow \RR$. The { \em Kolmogorov scaling function}  of $f$  is the function $\eta_f (p)$ implicitly defined by 
\begin{equation} \label{kolmo} \forall p >0, \hspace{8mm} 
\int |f(x+h) - f(x)|^p dx \quad\sim\quad   |h|^{\eta_f(p)},
\end{equation}
the symbol $\sim$ meaning that 
\BE  \label{scalKolmo}  \hspace{6mm} \eta_f (p) =\liminf_{|h |  \rightarrow 0} \frac{\log \left(\displaystyle\int |f(x+h) - f(x)|^p dx\right) }{\log |h| } . \EE 
Note that, if $f$ is smooth, then one has to use differences of order 2 or more   in order to define correctly the scaling function. 
Kolmogorov proposed to use this tool as a way to determine if some simple stochastic processes  are fitted to model the velocity of turbulent fluids at small scales, and  a first success of this approach was that  fractional Brownian motions (see Section \ref{Univariate2}) do not yield correct models (their scaling functions are linear, whereas the one measured on turbulent flows are significatively concave \cite{arneodoetal9600}). 

An important interpretation of the  Kolmogorov scaling function  can be given  in terms of { \em global smoothness}  indices  in families of functions spaces:    the spaces $\mbox{Lip}  ( s , L^p (\RR^d)) $ defined as follows. 
 Let $s  \in (0,1)$, and  $p \in [1, \infty]$; 
$ f \in \mbox{Lip}  ( s , L^p  (\RR^d)) $ if $ f \in L^p (\RR^d)$ and 
\begin{equation} \label{nicol1}  \exists C>0, \;\; \forall h  >0,  \hspace{6mm} 
\int |  f(x+h) -f(x) |^p  dx  \leq C |h|^{sp}    \end{equation}
(here also, larger smoothness indices $s$  are  reached by replacing  the first-order difference 
 $|  f(x+h) -f(x) |$ by  higher order differences).  
It follows from (\ref{kolmo}) and  (\ref{nicol1}) that, 
\begin{equation} \label{nicol}  \forall p \geq 1, \hspace{6mm}  \eta_f (p) = p \cdot \sup \{ s: f \in \mbox{Lip}  ( s , L^p  (\RR^d)) \} .
\end{equation}

An alternative formulation of the scaling function can be given in terms of global regularity indices supplied by Sobolev spaces, the definition of which we now  recall. 

\BD   \label{sobspac} 
 Let  $s \in \RR $ and  $p \geq 1$. A  function $f$ belongs to the  {  Sobolev space  } 
$L^{p,s} (\RR^d)$  if 
$\left( Id -\Delta  \right)^{s/2}
 f \in L^p$, where $ g= (Id -\Delta )^{s/2}f$  is  defined through its Fourier transform  as  
 \[ \hat{g}(\xi ) = ( 1 + | \xi |^2)^{s/2}\hat{f}(\xi ). \]
 \ED 
 
This  definition  amounts to state that the  fractional derivative of $f$  of order  $s$
 belongs to  $L^p$. The classical embeddings between the Sobolev and the  $\mbox{Lip}  ( s , L^p ) $  spaces  imply that 
\begin{equation} \label{nicol2}  \forall p \geq 1, \hspace{6mm}  \eta_f (p) = p \cdot \sup \{ s: f \in   L^{p,s} (\RR^d) \} .
\end{equation}
In other words, the scaling function  tells, for each $p$, the order of (fractional) derivation of $f$ up to which $f^{(s)}$ belongs to $L^p$.

A limitation of the use of the Kolmogorov scaling function for classification purposes is that many models display almost identical scaling functions (a  typical example is supplied by the velocity of  fully developed turbulence, see e.g. \cite{muzyetal91,LASHERMES:2008:A}); the next challenge therefore is to construct alternative scaling functions which would allow to draw distinctions between such models. 
A major advance in this direction was reached in  1985 when Uriel Frisch and Giorgio Parisi proposed another interpretation of the scaling function in terms of  { \em pointwise singularities} of the data \cite{ParFri85}.  In order to state their assertion, we first need the recall the  most commonly used notion of pointwise regularity. 

\BD \label{defholdponc} 
  Let $f: \; \RR^d  \rightarrow  \RR $  be a { locally   bounded function}, $x_0 \in   \RR^d$ and  let $\gamma \geq 0$; $f $ belongs to $ C^\gamma (x_0)$ if there exist $C>0$, $R >0$  and a polynomial  $P$ of degree less than  $\gamma$ such that  
\[ \mbox{ if} \;\; |x-x_0| \leq R,  \;\; \mbox{ then} \hspace{8mm}   |f(x) -P(x-x_0) | \leq C | x-x_0|^\gamma . \]  
The H\"older exponent of    $f$ at $x_0
$ is  
\BE \label{holexpo}   h_f  (x_0) =\sup \left\{ \gamma : \;\; f \;\; \mbox{ is  }   \;\;   C^{\gamma
} (x_0) \right\} . \EE 
\ED 
Some functions have a very simple H\"older exponent. For instance, 
the H\"older exponent of  the Weierstrass functions ${ \cal W}_{a, \ome}$ is  constant  and  equal to $\ome$ at every point (such functions are referred to as { \em monoh\"older functions}); since $\ome <1$ we thus recover the fact that ${ \cal W}_{a, \ome}$ is nowhere differentiable.  
However, the H\"older exponent of other functions turn out  to be extremely irregular, and  U. Frisch and G. Parisi introduced the { \em \hspace{-2mm}  multifractal spectrum }  ${ \cal D}_f $ as a  new quantity which allows to quantify  some of its properties: 
 ${ \cal D}_f (H)$  denotes the fractional dimension of the { \em \hspace{-2mm}   isoregularity sets}, i.e. the sets 
 \BE \label{isoh} \{ x : \quad h_f  (x) = H\} .  \EE 
 Based on   statistical physics  arguments, they  proposed  the following relationship between the scaling function  and ${ \cal D}_f (H)$: 
 \begin{equation}   \label{formul} { \cal D}_f (H)   =   \inf_{p  }  \left( d+Hp - \eta_f (p)  \right), \end{equation}
which is referred to as the { \em \hspace{-2mm}   multifractal formalism}, see \cite{ParFri85} (we will discuss in Section \ref{Univariate1}   the ``right'' notion of fractional dimension  needed here). 
 Though the remarkable intuition which lies behind this formula proved extremely fruitful,    it needs to be improved  in order to be completely effective; indeed
many natural processes  used in signal or image modelling do not follow this formula if one tries to extend it to negative values of $p$, see \cite{AJL05}; additionally, the only mathematical result relating the spectrum of singularities and the Kolmogorov scaling function in all generality  is  very partial, see \cite{jmf2,Jaf7}. In Section \ref{Univariate2} we will discuss  (\ref{formul}), and see how it needs to be reformulated  in terms of wavelet expansions in order to  reach a fairly general level of validity. In Section \ref{Poitexp} we will discuss the relevance of the H\"older exponent \eqref{holexpo} and introduce alternative exponents which are better fitted to the analysis of large classes of real-life data. Their characterization  requires the introduction of orthonormal wavelet bases. This tool and its relevance for global regularity is recalled in Section \ref{Univariate3} and the characterizations of pointwise regularity which they allow are performed in Section \ref{subsec:plead}. This leads to a classification of pointwise singularities which yields  a precise description of the oscillations of the function in the neighbourhood of its singularities which is developed in Section \ref{classif}. This implications of this classification on the different formulations of the multifractal formalism are developed in Section  \ref{mathresmf}.  The tools thus developed  are applied to marathon runners physiological data (heart rate, acceleration,   cadence, i.e. number of steps per minute)  in Section \ref{Marathon1}; 
thus showing   that they lead to a sharper analysis of the physiological modifications during the race. The numerical results derived on real-life data have been  obtained using the Wavelet $p$-Leader and Bootstrap based MultiFractal analysis (PLBMF) toolbox available on-line at  \\
$https://www.irit.fr/~Herwig.Wendt/software.html$ 

The explosion of data sciences recently made available  collections  of  signals the singularities of which are  expected to be related in some way; typical examples are supplied by EEG collected at different areas of the brain, or by collections of stock exchange prizes. The purpose of   Section \ref{Multivariate} is to address the extension of   multifractal analysis to the multivariate setting, i.e.  to several functions.   In such situations,  a pointwise regularity exponent $h_i(x)$ is associated with each signal $f_i (x)$  and the challenge is to recover the { \em joint multivariate spectrum} of the $f_i  $ which  is defined as the fractional dimension of 
the sets of points  $x$ where each of the exponents $h_i(x)$  takes a given value:  If $m$ signals are available, we define
\BE E_{f_1, \dots, f_m}  (H_1, \dots, H_m)  =  \{ x : \quad h_{1}  (x) = H_1, \dots, h_{m}  (x) = H_m  \} ,  \EE
and the  {\em joint  multifractal spectrum} is
\BE D_{f_1, \dots, f_m} ( H_1, \dots, H_m) = \dim (E_{f_1, \dots, f_m}  (H_1, \dots, H_m)  )  . \EE
These notions were  introduced by C.~Meneveau \emph{et al.}~in the seminal paper  \cite{Meneveau90}  
which addressed  the joint analysis of the dissipation rate of kinetic energy and passive scalar fluctuations for fully developed turbulence,
and a  general abstract setting  was proposed by J.~Peyri\`ere in \cite{pey2004}; In Section \ref{MultivariateSpec}, we introduce the mathematical concepts which are relevant to this study.  In Section \ref{MultivariateWSF} we give a probabilistic interpretation of the  scaling functions introduced in Section \ref{Univariate}, and we show how they naturally lead to a 2-variable extension in terms of correlations.   The initial  formulation of the multifractal formalisms  based on extensions of the Kolmogorov scaling function suffers from the same drawbacks as in the univariate  case.  This leads naturally to a reformulation of the  multifractal formalism which is  examined in Section \ref{MultivariateMF}, where we also investigate the additional advantagess supplied by multivariate  multifractal analysis  for singularity classifications.   In order to investigate its relevance, we study a toy-example which is supplied by Brownian motions in multifractal time in Section \ref{BrwonianMT}.    In Section \ref{MultivariateMar}, we  illustrate the mathematical results thus collected  by  applications to the joint analysis  of heartbeat, cadence and acceleration of marathon runners.

\section{Univariate multifractal analysis} 

\label{Univariate}

\subsection{  The multifractal spectrum} 

\label{Univariate1}

In order to illustrate the motivations of multifractal analysis, let us come back to the initial problem we mentioned: How badly  can a continuous function behave? We mentioned the surprising   result of  Mazurkiewicz   and Banach  stating that a generic  continuous function is  nowhere differentiable, and the Weierstrass functions yield examples of continuous functions  which may have an arbitrarily small (and constant) H\"older exponent.  This can actually be improved:  A generic continuous function satisfies 
\BE \label{genercont} \forall x  \in \RR, \qquad h_f (x) = 0, \EE
see \cite{HSY}: At every point the  H\"older exponent  of $f$ is  as bad as possible.  An example of such  a continuous function is supplied by  a slight variant of Weierstrass functions:
\[ f(x) = \sum_{j=1}^\infty \frac{1}{j^2} \sin (2^j x).  \]
  Let us now consider a different functional setting:  Let  $f : [0,1] \longrightarrow [0, 1]$ be an  increasing function.  At any given point $x\in [0,1]$  $f$ can have a discontinuity at $x$, in which case  $h_f (x) = 0$. Nonetheless, this worse possible behavior cannot be met everywhere: An important theorem of Lebesgue states that  $f$ is almost everywhere differentiable and therefore satisfies
  \[  \mbox{ for almost every } x \in [0, 1], \qquad  h_f (x) \geq 1. \] 
  The global regularity assumption (the fact that $f$ is increasing implies that its derivative in the sense of distributions  is a bounded Radon measure) implies that, in sharp contradistinction with  generic continuous functions, the set of points such that $h_f (x) <1$ is ``small'' (its Lebesgue measure vanishes). 
 On other hand, the set of points where it is discontinuous   can be an arbitrary  countable set (but one easily checks that it cannot be larger).  What can we say about the size  the sets of points with intermediate regularity (i.e. having H\"older exponents between 0 and 1), beyond the fact that they have a vanishing Lebesgue measure? 
Answering this problem requires to use some appropriate notion of ``size''  which  allows to draw differences between sets of vanishing Lebesgue measure.  The right mathematical notion fitted to this problem   can be  guessed using  the following argument. 
Let 
\[ \EAL = \{ x: f \notin C^\al (x) \} . \]
Clearly, if  $x \in \EAL$, then there exists  a sequence of dyadic intervals  
\BE  \label{dyadint} \la_{j,k} =  \left[\frac{k}{2^j},\frac{k+1}{2^j}  \right] \EE  such that 
\begin{itemize}
\item $x$  belongs either to  $\la_{j,k}$  or to one of its two closest neighbours  of the same width,
\item  the increment of  $f$ on  $\la_{j,k}$ is larger than $2^{-\al j} = |\la_{j,k}|^\al$ (where $|A|$ stands for the diameter of the set $A$). 
\end{itemize}
Let  $\ep >0$, and consider the { \em maximal} dyadic intervals  of this type of width less than   $\ep /3$,  for all possible  $x \in \EAL$, and denote this set  by $\La^\ep_\al$. 
 These  intervals are  disjoint (indeed two dyadic intervals are either disjoint or one is included in the other); and, since  $f$ is increasing,  the increment of  $f$ on $[0,1]$ is bounded by the sum of the increments on these intervals.  Therefore
\[ \sum_{ \la \in \La^\ep_\al} | \la |^\al \leq f(1)-f(0). \]
The intervals  $3 \la$ (which consists in the dyadic interval $\la$ and its two closest neighbours of the same length)  for $ \la \in \La^\ep_\al$ form an  { \em $\ep$-covering }\ of  $\EAL$ (i.e. a covering by intervals of length  at most $\ep$), and this  $\ep$-covering satisfies 
\[ \sum_{ \la \in \La^\ep_\al} | 3\la |^\al =  3^\al  \sum_{ \la \in \La^\ep_\al} | \la |^\al \leq f(1)-f(0). \]
This property can be interpreted as stating that the  { \em $\al$-dimensional Hausdorff measure}  of $\EAL$ is finite; we now give  a precise definition of this notion.

\BD  
\label{defmeshaus} 
Let   $A$ be a subset of $ \RR^d$. 
If $\ep>0$  and  $\delta \in [0,d]$,  let
\[ M^\delta_{\ep} =\inf_R \;  \left( \sum_{  i} | A_i |^\delta \right) ,\]
where $R$ is an  $\ep$-covering of   $A$, i.e. a covering of  $A$  by bounded sets  $\{ A_i\}_{i \in
\NN}$
 of diameters  $| A_i | \leq \ep$
(the infimum is therefore taken on all  $\ep$-coverings).
For any $\delta \in [0,d]$, the 
$\delta$-dimensional  Hausdorff measure of 
$A$ is 
\[  mes_\delta (A) = \displaystyle\lim_{\ep\rightarrow 0}  M^\delta_{\ep}. \]
\ED  

One can show that there exists $\delta_0 \in [0,d]$ such that 
\[ \left\{ \begin{array}{l} \forall \delta < \delta_0, \;\;\; 
 mes_\delta (A) = + \infty  \\ \forall \delta > \delta_0, \;\;\; 
 mes_\delta (A) = 0
. \end{array}\right. \]
This critical  $\delta_0$ is called the   { \em Hausdorff dimension}  of  $A$, and is  denoted by $\dim (A)$ (and an important convention  is that, if   $A$ is empty,  then $\dim \,(\emptyset )= -\infty $).

The example we just worked out shows that  a global regularity information on a function   yields information on the Hausdorff dimensions of its  sets of H\"older singularities. This indicates that the Hausdorff dimension is the natural choice in (\ref{formul}), and  motivates the following definition.

\BD 
Let  $f: \RR^d \rightarrow \RR$ be a   locally  bounded function. 
The  { multifractal H\"older spectrum } of $f$ is the function  
\[  { \cal D}_f (H)= \dim (\{ x: \hspace{3mm}  h_f(x) = H\}   ), \] where  $\dim$ denotes the Hausdorff dimension.
\ED

This definition justifies the denomination of { \em multifractal functions}:   One typically considers  functions $f$ that have non-empty isoregularity sets  \eqref{isoh} for $H$ taking all values in an interval of positive length, and therefore one   deals  with  an   infinite  number of fractal sets $E_f (H) $.  The result we obtained thus implies  that, if $f$ is  an increasing function, then 
\BE \label{majspecinc} { \cal D}_f (H) \leq H. \EE   This  can  be reformulated  in a  function space setting which puts in light  the sharp contrast with   (\ref{genercont}): Indeed,  recall that   any function of bounded variation is the difference of an increasing and a decreasing function; we have thus obtained the following result.  

\BP Let $f: \RR \rightarrow \RR$ be a function of bounded variation. Then its multifractal spectrum satisfies 
\[ \forall H, \qquad { \cal D}_f (H) \leq H . \]
\EP 


{ \bf Remark:} This result does not extend to several variables functions of bounded variation which, in general, are not locally bounded, in which case their H\"older exponent is not even well defined. 

\subsection{  Alternative  formulations of the  multifractal formalism} 

\label{Univariate2} 

We mentioned that (\ref{formul}) yields a poor estimate of the multifractal spectrum. A typical example is supplied by sample paths of  { \em  fractional Brownian motion}  (referred to as fBm), a family of stochastic processes introduced by Kolmogorov \cite{Kolmo40}, the  importance of which   was put in light for  modeling by Mandelbrot  and Van Ness  \cite{mvn68}. This family is indexed by a  parameter $\al\in (0,1)$, and generalizes Brownian motion (which corresponds to the case $\al= 1/2$); fBm   of index $\al $  is the only centered 
Gaussian random process $B^\al $  defined on $\RR^+$ which  satisfies 
\[ \forall x, y \geq 0\qquad \EEE ( | B^\al (x)-B^\al (y)|^2) = |x-y|^{2\al} .\]
  FBm plays an important  role in signal processing  because it
supplies the most simple one parameter family of stochastic processes with stationary
increments.  Its sample paths  are monoh\"older  and  satisfy 
\[ \mbox{   a.s.  }   \forall x , \qquad h_{B^\al} (x) =\al ,  \] 
(see \cite{KahCamb} (and \cite{EssLoos} for a recent sharp analysis of the pointwise regularity of their sample paths) so that their  multifractal spectrum is 
\[ \mbox{ a.s. }  \quad \forall H, \qquad 
\left\{\begin{array}{rll}
 { \cal D}_{B^\al} (H) =   & 1 & \quad\mbox{if }  H  = \al\\[2mm]
= & -\infty & \quad\mbox{else. } 
\end{array}\right.
\]
However, the right hand-side of (\ref{formul}) yields  a different value for $H \in (\al, \al +1]$: It coincides almost surely with the function  defined by
\[
\left\{\begin{array}{rll}
 { \cal L}_{B^\al} (H) = & \al +1 - H & \quad\mbox{if } H \in [\al,  \al +1] \\[2mm]
= & -\infty & \quad\mbox{else, } 
\end{array}\right.
\]
see \cite{JAFFARD:2010:A,JAFFARD:2006:A,MandMemor}.  
This is due to the fact that the  decreasing part of the spectrum is recovered from negative values of $p$ in (\ref{formul}), and the corresponding  integral is not well defined for negative $p$s, and may even diverge.  It follows that sharper estimates of the multifractal spectrum require a renormalization procedure which would yield a numerically robust output for negative $p$s.  Several methods have been proposed to solve this deadlock.  They are all based on a modification of the Kolmogorov scaling function in order to incorporate  the  underlying intuition that it should include some pointwise regularity information.  A consequence will be that they  provide an extension of the scaling function to negative $p$s. This extra range of parameters plays a crucial role in several applications where it is required for classifications, see e.g. \cite{muzyetal91,LashermesEuropeanPhysicalJournalB2008} where  the validation of turbulence models is considered, and for which the key values of the scaling function which are needed to draw significative differences  between these models are obtained for $p<0$.

A first method is based on the  { \em continuous wavelet transform}, which is defined as follows. 
Let  $\psi$ be a { \em wavelet}, i.e. a well localized, smooth function with, at least, one vanishing moment.  The continuous wavelet transform of  a one-variable function $f$ is 
\BE   \label{dec} C_{a,b} (f) = \frac{1}{a}\int_\RR f(t) \psi \left(\frac{t-b}{a}\right)  dt \qquad ( a >0, \quad b \in \RR );\EE

Alain Arneodo, Emmanuel Bacry and  Jean-François Muzy proposed to replace, in the integral (\ref{kolmo}),
the increments    $ | f (x+\delta) -f (x) |$ at scale $\delta$  by the continuous wavelet transform 
  $C_{a,b} (f) $ for $a = \delta $ and $b = x$.  This choice  follows  the heuristic that  the  continuous wavelet transform satisfies
  $   |C_{a,b} (f)|   \sim  a^{h_f (x)} $ 
  when $a$ is small enough and  $| b-x| \sim a $. Note that it is not valid in all generality, but typically  fails for { \em oscillating singularities}, such as the chirps (\ref{chirp}).  Nonetheless Yves Meyer showed that this heuristic actually characterizes another pointwise  regularity exponent, the { \em weak scaling exponent}, see \cite{MeyWVS}. Assuming that the data do not include oscillating singularities, 
   the  integral (\ref{kolmo})  is discretized  and replaced  by the more meaningful values of the continuous wavelet transform i.e.  at its local maxima \cite{ABM95}; if we denote by  $b_k $ the points where these extrema are reached at the scale  $a$, 
 the integral (\ref{kolmo}) is thus replaced by the sum 
  \BE  \label{foncechbis}  \sum_{b_k} | C_{a,b_k} (f)|^p   \sim  a^{  \zeta_f (p)}  \quad \mbox{ when } a \rightarrow 0  , \EE
  This reformulations using the { \em multiresolution quantities } $| C_{a,b_k} (f)|$ yields better numerical results than when using  the  increments $ | f (x+\delta) -f (x) |$;  above all, the restriction to the local suprema  is a way to bypass the small values of the increments  which were the cause of the divergence of the integral  (\ref{kolmo}) when $p$ is negative. Numerical experiments consistently show that the multifractal formalism based on these quantities yields the correct spectrum for the fBm, and also for large collections of mutifractal models, see \cite{AADMV}. 
  
  Another way  to obtain a numerically robust procedure in order  to perform multifractal analysis is supplied by  
{ \em \hspace{-2mm} Detrended Fluctuation Analysis}  (DFA) : From the definition of the H\"older exponent, 
Kantelhardt et al. \cite{kantelhardt2002multifractal}   proposed  the following   multiresolution quantity based on  the following local $L^2$ norms
\begin{equation}
\label{equ:mradfa}
\TMFD(a,k) =  \left( \frac{1}{a} \sum_{i=1}^{a} |f(ak+i) - P_{k,a, N_P}(i) |^2 \right)^{\frac{1}{2}}, k =1, \ldots, n/a,
 \end{equation}
 where  $n $ denotes the number of available samples and $P_{t,a,N_P}$ is a polynomial of degree $N_P$  obtained by {local fit to $f$} on portions of length proportional to $a$. 
The integral (\ref{kolmo}) is now replaced by
$$
S_{mfd}(a,q)=  \frac{a}{n}  \sum_k^{n/a} \TMFD(a,k)^q  \sim a^{\zeta_{mfd}(q) },
$$
 and the multifractal spectrum is  obtained as usual through a Legendre transform of this new scaling function $\zeta_{mfd}$, thus  yielding  the  \emph{multifractal detrended fluctuation analysis} (MFDFA).  Note that, here again, we cannot expect the  multifractal formalism based on such a formula to be fitted to the H\"older exponent: The choice of an $L^2$ norm in (\ref{equ:mradfa})  is rather adapted to an alternative pointwise exponent, the  2-exponent, which is defined through local $L^2$-norms, see Def. \ref{def:pexp} (and \cite{PART2} for an explanation of this interpretation). 
The MFDFA formalism  performs  satisfactorily and is commonly used in applications (cf., e.g., \cite{galaska2008comparison,Wang2012}). 

The methods we mentioned  meet the following limitations:  They cannot be taylored to a particular pointwise exponent: We saw that the WTMM is fitted to the weak-scaling exponent, and the MFDFA to the 2-exponent. They lack of   theoretical foundation, and therefore the estimates that they yield on the multifractal spectrum   are not backed by mathematical results.  In practice, they are difficult to extend to data in two or more variables (for MFDFA, the computation of local best fit polynomials is an intricate  issue).   The obtention of an alternative formulation of the multifractal formalism  which  brings an answer to these two  problems requires a detour through the  notions of pointwise  exponents, and their characterizations.  

\subsection{Pointwise exponents} 

\label{Poitexp}

At this point we need to discuss the different notions of pointwise regularity. One of the reasons is that, though H\"older regularity is by far the one which is most used in mathematics and in applications,  
it suffers a major limitation: Definition \ref{defholdponc}  requires $f$ to be locally bounded. In applications,  this limitation makes the H\"older exponent  unfitted  in many  settings where modelling data by locally bounded functions is inadequate; in Section \ref{Univariate3}  we will give a numerically simple criterium which allows to verify if  this assumption is valid, and   we will see that the physiological data we analyse are typical examples  for which it is not satisfied.   On the mathematical side too, this notion often is  not relevant.  A typical example is supplied  by the Riemann series defined as 
\BE \label{riem}  \forall x \in \RR, \qquad {\mathcal R}_s (x) = \sum_{n=1}^\infty \frac{\sin(n^2 x) }{n^s} ,\EE
which, for $s > 1$, are locally bounded and turn out to be multifractal (in which case their multifractal analysis can be performed using the H\"older exponent \cite{Vindas,JaffRie}), but it is  no more the case if $s <1$, in which case  an alternative analysis is developed  in \cite{SeurUb} (using the $p$-exponent for $p =2$, see Def. \ref{def:pexp} below).

There exist two ways to deal with such situations. 
The first one consists in  first regularizing  the data, and then analyzing the new data thus obtained. Mathematically, this means that a { \em fractional integral} is performed on the data.   Recall that, if $f$ is a tempered distribution defined on $\RR$, then the fractional integral of order $t$ of $f$, denoted by $   f^{(-t )}$  is defined as follows: 
 Let $(Id-\Delta
)^{-t /2}$ be the convolution operator which amounts to
multiplying the Fourier transform of $f$  with 
$(1+|\xi |^2)^{-t /2}$. The fractional integral of order $t$ of $f$ is the function  
\[    f^{(-t )} =(Id-\Delta )^{-t /2} ( f) . \] 
If $f$ is large enough, then $f^{(-t )}$ is a locally bounded function, and one can consider the H\"older exponent of $t$ (the exact condition under which this is true is that $t$ has to be larger than the exponent $\Hmin$ defined below   by  \eqref{hmin1} or equivalently by  \eqref{caracbeswav3hol}).  This procedure 
 presents the obvious disadvantage of not yielding  a direct analysis of the data  but of a smoothed version of them.

The other alternative available in order to characterize the pointwise regularity of non-locally bounded functions  consists in  using a weaker notion of pointwise regularity, the $p$-exponent, which we now recall. We define $B(x_0, r)$ as the ball of center $x_0$ and radius $r$.

\BD  \label{def:pexp} 
 Let $p \geq 1$ and assume that $f\in  L^p_{loc} (\RR^d)$.  
Let  $\alpha \in \RR$;  
 $ f $   belongs to $T^p_\alpha (x_0)$  if there exists  a constant $ C$  and a polynomial $P_{x_0}$  of degree  less than $\alpha $  such that, for $r$ small enough,  
\BE 
\label{pexpa} 
 \left( \frac{1}{r^d} \int_{B(x_0, r} | f(x) -P_{x_0}(x)|^p dx  \right)^{1/p} \leq C r^\alpha.
\EE 
  The { \bf $p$-exponent}   of  $f$ at $x_0$ is 
\BE  
\label{equ-pexp} 
h_f^p(x_0) = \sup \{ \alpha : f \in T^p_\alpha (x_0)\}
\EE
(the case $p= + \infty$ corresponds to the H\"older exponent). 
 \ED
 
  This definition was introduced by      Calder\'on and Zygmund   in 1961 in order to obtain pointwise regularity results for the solutions of elliptic PDEs,  see \cite{CalZyg}. For our concern, it has the important property of being well defined under  the assumption that $f \in L^p_{loc}$. For instance, in the case of the Riemann series (\ref{riem}), an immediate computation yields that they belong to $L^2$ if $s >1/2$ so that, if $ 1/2 < s <1$,  $p$-exponents with $p \leq 2$  are relevant to study their regularity, in contradistinction with  the H\"older exponent  which won't be defined. Another example of multifractal function which is not locally bounded is supplied by Brjuno's function, which plays an important role in holomorphic dynamical systems, see \cite{MaMoYo}.  Though its is nowhere locally bounded, it belongs to all $L^p$ spaces and its multifractal analysis using $p$-exponents has been performed in \cite{JaMar}. 
 Note that  $p$-exponents can take values  down to $-d/p$,  see \cite{PART1}. Therefore, they allow the use of { \em negative regularity exponents}, such as singularities  of the form $f(x) =  1/|x-x_0|^\alpha$ for $\alpha<d/p$.
 
 The general framework supplied by multifractal analysis now is ubiquitous in mathematical analysis and has been successfully used in a large variety of mathematical situations, using diverse  notion of pointwise exponents such as pointwise regularity of probability measures  \cite{Bmp92}, rates of convergence or divergence of series of functions (either trigonometric \cite{Aubry06,BaHeu} or wavelet  \cite{Aubry06,EssJaff})  order of magnitude of ergodic averages \cite{Ling,Ling2}, to mention but a few.

\subsection{  Orthonormal wavelet decompositions} 

\label{Univariate3}

Methods based on the use of orthonormal wavelet bases  follow the same motivations we previously developed, namely  to  construct alternative scaling functions based on multiresolution quantities which ``incorporate'' some pointwise regularity information.  However,   we will see that  they allow to turn some of the limitations met by the  previously listed methods, and they enjoy   the following  additional properties:
\begin{itemize}
\item numerical simplicity, 
\item  explicit links with  pointwise exponents (which, as we saw, may differ from the H\"older exponent),
\item no need to construct local polynomial approximations (which is the case for  DFA methods now in use),
\item mathematical results hold concerning  either  the validity of the multifractal formalism supplied by (\ref{formul}) or of some appropriate extensions; such results  can  be  valid for all functions, or   for ``generic'' functions, in the sense of Baire categories, or for  other notions of genericity.  
\end{itemize}
Let us however mention an alternative technique which was proposed  in \cite{Abel} where multiresolution quantities based on local oscillations, such as 
\[  d_\la = \sup_{ 3 \la} f(x) -\inf_{3\la} f(x) ,  \] or higher order differences such as 
\[  d_\la  = \sup_{x,y \in 3 \la} \left|  f(x) + f(y) -2  f\left(\frac{x + y}{2}\right)\right| ,  \]
 and which wouldn't present the third problem that we mention. However, as far as we know, they haven't been tested numerically. 

One of the reasons for these remarkable properties is that  (in contradistinction with  other expansions, such as e.g. Fourier series)
 wavelet  analysis   allows to characterize both global   and pointwise regularity  by simple conditions on the moduli of the wavelet coefficients;  as already mentioned, the multifractal formalism raises the question of how global and pointwise regularity are interconnected;   wavelet analysis therefore is a  natural tool in order to investigate this question and  this explains why it  was at the origin of  major   advances in multifractal  analysis both in theory and applications.

 We now recall the definition of orthonormal wavelet bases. For the sake of  notational simplicity, we assume in all the remaining of Section \ref{Univariate}  that $d=1$, i.e.  the functions we consider are defined on $\RR$, extensions in several variables  being  straightforward.  Let $\varphi(x) $ denote  an  smooth function with fast  decay, and good joint time-frequency localization, referred to as the {\em \hspace{-2mm} scaling function},   and let
 $\psi(x) $ denote  an   oscillating function (with $N$ first vanishing moments), with fast  decay, and good joint time-frequency localization, referred to as the {\em \hspace{-2mm} wavelet}.  
These functions  can   be chosen such that the 
\BE
\label{wavbasbad1}  
\varphi (x-k), \hspace{5mm} \mbox{ for} , \hspace{3mm}  k\in \ZZ
\EE
and 
\BE
\label{wavbasbad}  
2^{j/2}\psi (2^{j}x-k), \hspace{5mm} \mbox{ for} , \hspace{3mm}   j \geq 0, \;\; k\in \ZZ
\EE
form an orthonormal basis of $L^2 (\RR)$ \cite{Mey90I}.
The  wavelet coefficients  of  a function $f$ are defined as
\begin{equation}
\label{eq-wc}  c_{ k} =   \displaystyle\int_{\R} f(x)  \;  \varphi (x-k) \, dx \qquad \mbox {and}  \qquad 
c_{ j,k} =   2^{j} \displaystyle\int_{\R} f(x)  \;  \psi (2^{j}x-k) \, dx
\end{equation}
Note the   use of an  $L^1$ normalization for the wavelet coefficients that better fits local regularity analysis.
 
 As stated above,
 the H\"older exponent  can be used as a measurement of pointwise regularity  in the  locally bounded functions setting only, see \cite{JAFFARD:2010:A}.  
Whether empirical data can be well-modelled by  locally bounded functions or not can be determined numerically through the computation of the {\em uniform H\"older exponent}  
$\Hmin$, which, as for the scaling function, enjoys a function space characterization
\BE \label{hmin1} \Hmin = \sup\{ \al: f \in C^\al (\RR )\}, \EE 
where $C^\al (\RR )$ denotes the usual H\"older spaces. Assuming that $\varphi$ and $\psi$ are smooth enough and that $\psi$  has enough vanishing moments, then the exponent $\Hmin$
  has the following simple wavelet characterization:
 \BE 
\label{caracbeswav3hol}  
\Hmin =\liminf_{j \rightarrow + \infty} \;\;  
 \;\;  \frac{ \log 
\left(  \displaystyle  \sup_{  k}  | c_{ j,k}  |   \right) }{\log (2^{-j})}. 
\EE
It follows that, if {  $\Hmin > 0$}, then $f$ is  a continuous  function,  whereas, if  {  $\Hmin < 0$}, then $f$ is  not a locally bounded function,  see \cite{MandMemor,Bergou}.

In numerous real world applications the restriction $\Hmin > 0$  constitutes a severe limitation; we will meet such examples in the case of physiological data (see also  \cite{MandMemor} for other examples).
From a pratical point of view,  the regularity of the wavelets should be larger than  $\Hmin$ in order  to compute the estimation of $\Hmin$. In the applications that we will see later, we took Daubechies compactly supported wavelets of increasing regularity and we stopped as soon as we found a threshold beyond which there is no more modification of the results. In our case, we stopped at order 3.
In applications, the role of $\Hmin$ is twofold: It can be used as a classification parameter and it tells whether a multifractal analysis based on the H\"older exponent is licit.  Unlike other multifractality parameters that will be introduced in the following, its computation does   not require  a priori assumptions: It can be defined in the widest possible setting of tempered distributions.  \\ 
We represent these two types of data on Fig. \ref{hmin} for a marathon runner. The race is composed of several stages including a warm-up at the beginning, a recovery at the end of the marathon, and several moments of small breaks during the marathon. The signal was cleaned by removing the data that did not correspond to the actual race period (warm-ups, recoveries and breaks) and by making continuous connections to keep only the homogeneous parts. This type of connection is suitable for regularities exponents lower than 1 as in the case of our applications.

\begin{figure}
    \centering
    \includegraphics[scale=0.3]{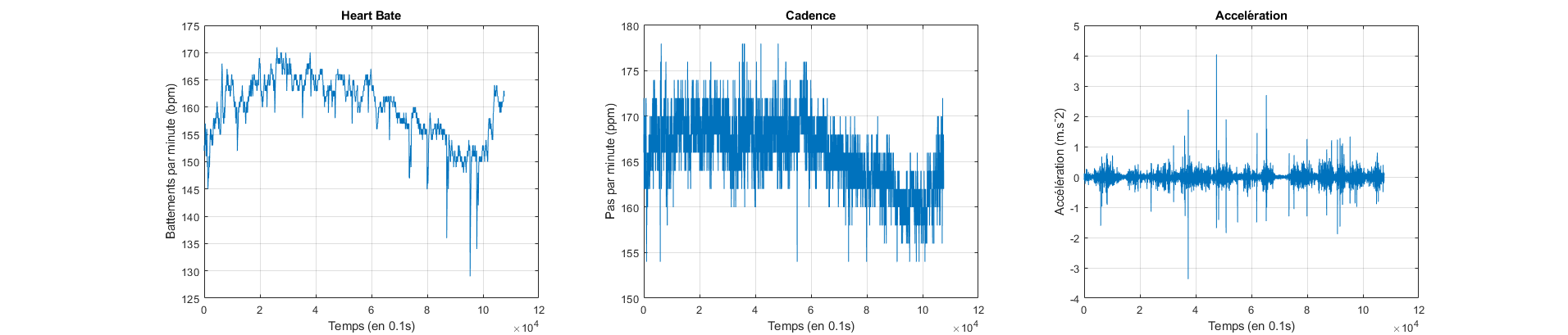}
    \caption{Representation of data: heart rate (left) in beats per minute, cadence (middle) in steps per minute and acceleration (top) in meters per second squared. The time scale is in 0.1s}
    \label{data}
\end{figure}

\begin{figure}
    \centering
    \includegraphics[scale=0.3]{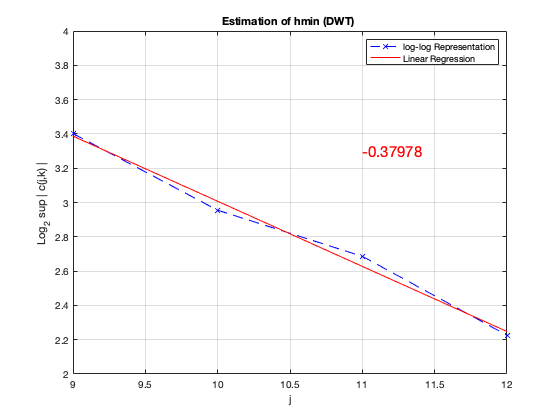} \quad
    \includegraphics[scale=0.3]{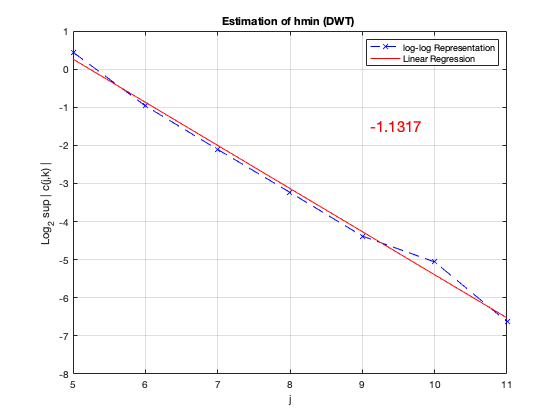}
    \caption{Estimation by log-log regression of the $h_{min}$ of a heart rate (left) and an acceleration (right). The points of the regression
    line up successfully along a close to straight line
    thus showing that   the values of $h_{min}$, are precisely estimated and  are negative. It follows that a   multifractal analysis based on H\"older exponent cannot be performed on these data.}
    \label{hmin}
\end{figure}

If $\Hmin <0$, then   a multifractal analysis based on the H\"older exponent cannot be developed, and the question whether  a multifractal analysis based on the  $p$-exponent    can be raised. Wavelet coefficients can also be used to determine  whether $f$ locally belongs to $L^p$ or not (which is the a priori requirement needed in order to use the corresponding $p$-exponent), see \cite{Bergou,Abel,MandMemor}:
Indeed, a simple wavelet criterium can be applied to check this assumption, through the computation of the   { \em wavelet structure function}. Let 
\begin{equation}
\label{equ-WSF} 
S_c(j,p) = 2^{-j} \displaystyle\sum_{ k }  | c_{ j,k}  |^p. 
\end{equation}
 The \emph{wavelet scaling function} is defined as
\BE \label{defscalond}  \forall p >0, \hspace{6mm} 
 \eta_f (p) =   \displaystyle \liminf_{j \rightarrow + \infty} \;\; \frac{\log \left( S_c(j,p)  \right) }{\log (2^{-j})}; \EE 
one can show that it coincides with the Kolmogorov scaling function if $p >1$, see \cite{jmf2}. 
The following simple criterion can be applied  in order to check if data locally belong to $L^p$ \cite{PART1}: 
\begin{equation}
\label{equ:etap}
\left. \begin{array}{rl}  \mbox{ if} \; \eta_f(p) >0 & \mbox{then} \; f \in L^p_{loc}, \\ 
\mbox{ if} \; \eta_f (p) < 0 & \mbox{then} \; f \notin L^p_{loc} . \\ 
\end{array}\right\}
\end{equation}

{ \bf Remarks:} The wavelet scaling function   enjoys the same property as $\Hmin$: Its computation does not  require some a priori assumptions on the data, and it can be defined in the general setting of tempered distributions.   Note that it is also defined for $p \in (0,1]$; in that case  the Sobolev space interpretation  of the scaling function has to be slightly modified: In Def. \ref{sobspac}  the Lebesgue space $L^p$ has to be replaced by the real  Hardy spaces $H^p$, see \cite{Mey90I} for the notion of Hardy spaces and their wavelet characterization. 
Note that these function space interpretations imply that the wavelet scaling function does not depend on the specific (smooth enough) wavelet basis which is used; it also implies that it is unaltered by the addition of a smooth function, or by a smooth change of variables, see \cite{Abel} and ref. therein. For the same reasons, these  properties also hold for the exponent  $\Hmin$; they  are required in order to derive intrinsic parameters for signal or image classification. In the following, we shall refer to them as { \em robustness properties}. In applications \eqref{defscalond}  can be used only if  $\eta_f(p)$ can be determined by a log-log plot regression, i.e.  when the  limit  actually is a limit, see e.g. Fig. \ref{scaling function for p=1}.  This means that the structure functions \eqref{equ-WSF} satisfy $S_c (j,p) \sim 2^{-\eta_f (p)j} $ in the limit of small scales, a phenomenon coined { \em scale invariance}.  The practical relevance  of the wavelet scaling function (and other multifractal parameters that we will meet later), comes from the fact that it can be used for classification of signals and images without assuming that the data follow an a priori model.

 \begin{figure}
    \centering
    \includegraphics[scale=0.3]{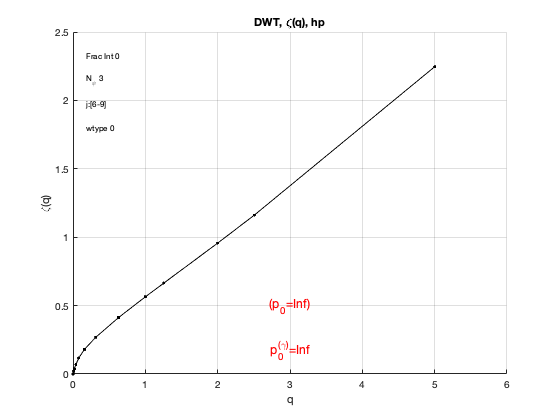} \quad
    \includegraphics[scale=0.3]{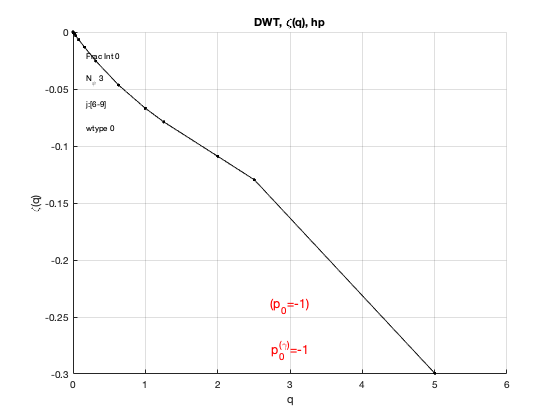}
    \caption{Wavelet scaling function of heart rate (left) and cadence (right) of a marathon runner. It allows to determine the values of $p$ such that $\eta_f (p)>0$. We conclude that  a multifractal analysis based on $p$-exponents  is directly possible for heart rate data, but not for the cadence, where the analysis will have to be carried out on a  fractional integral of the data}
    \label{scaling function}
\end{figure}

 \begin{figure}
    \centering
    \includegraphics[scale=0.3]{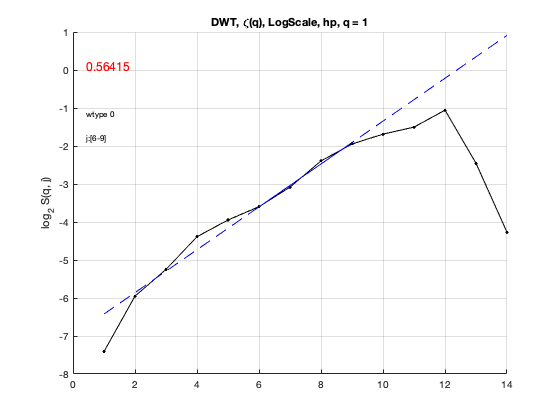} \quad
    \includegraphics[scale=0.3]{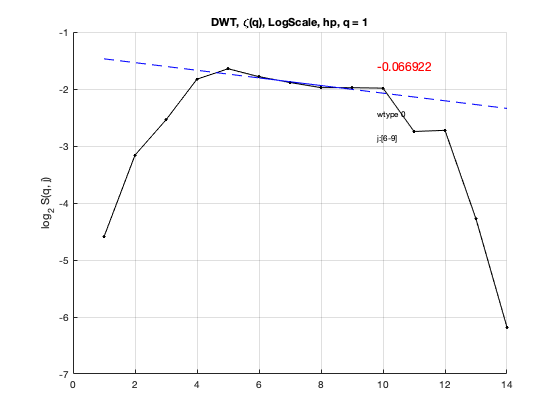}
    \caption{Estimation by log-log regression of the wavelet scaling function of heart rate (left) and cadence (right) for $p=1$. The slope of the regression is positive for heart rate and negative for cadence. These regressions, estimated for a suffiently large number of values of $p$ allow to plot the wavelet scaling functions, as shown in Fig. \ref{scaling function}}
    \label{scaling function for p=1}
\end{figure}

\subsection{Wavelet pointwise regularity characterizations}
 \label{subsec:plead}
 

One advantage of orthonormal wavelet based methods is that they allow to construct a multifractal analysis which is taylored for  a given $p$-exponent, which is not the case of the alternative methods we mentioned. We shall see in Sections \ref{classif} and \ref{Marathon1}
  the benefits of this extra flexibility. For this purpose, we have to construct multiresolution quantities (i.e., in this context,  a non-negative function defined on the collection of dyadic cubes) which are fitted to $p$-exponents. 
 We start by introducing more adapted notations for wavelets and wavelet coefficients; 
instead of the two indices $(j,k)$, we will use dyadic intervals \eqref{dyadint} 
and, accordingly, $c_{\lambda} =  c_{ j,k}   $, and $\psi_{\lambda} = \psi_{ j,k}$. 
The wavelet  characterization of  $p$-exponents requires the  definition of  {$p$-leaders}. 
If $f \in L^p_{loc} (\RR)$,  the  wavelet $p$-{\em leaders}   of $f$ are defined as

\begin{equation} 
\label{pleaders} 
 \pL_{j,k}   \equiv \pL_\la= \left(   \sum_{ \lambda' \subset 3 \lambda}  | c_{\lambda'} |^p \, \, 2^{j-j'} \right)^{1/p},
\end{equation}
where $j'\geq j$ is the scale associated with the sub-cube $\la'$ included in  $3 \lambda$  (i.e. $\la'$ has width $2^{-j'}$).
Note that, when $p = +\infty $ (and thus $f \in  L^{\infty}_{loc} (\RR)$), {$p$-leaders} boil down to  { \em  wavelet leaders}
\[ \ell_\lambda = \sup_{  \lambda ' \subset 3 \lambda}  | c_{\lambda'}|, \]
 \cite{Jaffard2004,Wendt:2007:E}.  
 
 Let us indicate where such quantities come from.  They are motivated by constructing quantities  based on simple conditions on wavelet coefficients and which  well approximate the local $L^p$ norm of Definition \ref{def:pexp}.  For that purpose we use the wavelet characterization of the Besov space $B^{0,p}_p$  which is ``close'' to $L^p$ (indeed the classical embeddings between  Besov and $L^p$ spaces imply that $B^{0,1}_p \hookrightarrow L^p \hookrightarrow B^{0,\infty}_p $); with the normalization we chose for wavelet coefficients, the wavelet characterization of $B^{0,p}_p$  is given by
 \[  f \in B^{0,p}_p \quad \mbox{ if} \qquad  \sum_k | c_k |^p < \infty  \qquad \mbox{and}  \qquad \sum_{j,k}  2^{(sp-1)j}| c_{j,k} |^p < \infty , \] 
 see \cite{Mey90I} and, because of the localization of the wavelets,  the restriction of the second sum  to the dyadic cubes $\la' \subset 3 \la$  yields an approximation of the local $L^p$ norm of $f-P$  around the interval $\la$ (the substraction of the polynomial $P$ comes from the fact that  the wavelets have vanishing moments so that  $P$ is reconstructed  by the first sum in  \eqref{wavbasbad1}, and the wavelet coefficients $c_{j,k}$ of $f$ and $f-P$ coincide).  Actually, the  uniform regularity assumption $\eta_f (p) >0$ (which we will make) implies that the quantities \eqref{pleaders} are finite. 
 
 Denote by $\lambda_{j,k} (x)$ the unique dyadic interval of length $2^{-j}$  which includes $x$; a key result is that both the H\"older  exponent and the $p$-exponent can be recovered from, respectively, wavelet leaders and $p$-leaders, according to the following formula.
 
 \BD \label{derived} Let $h(x)$ be a pointwise exponent and $(d_\la)$ a multiresolution quantity indexed by the dyadic cubes.  The exponent $h$ is derived from the $(d_\la)$ if 
 \begin{equation} 
\label{carachqf}   \forall x, \qquad 
h(x)  =  \liminf_{j \rightarrow + \infty}  \frac{ \log  \left(  d_{\lambda_{j,k} (x)} \right)  }{\log (2^{-j})}. 
\end{equation}
\ED 
 
It is proved in \cite{JaffToul, Bergou,JaffMel} that 
if  $\eta_f (p) >0$, then the   $p$-exponent  is derived  from $p$-leaders,  
and, if $\Hmin >0$, then the H\"older  exponent  is derived  from  wavelet leaders.  Note that  the notion of $p$-exponent can be extended to values of $p$ smaller that 1, see   \cite{JaffCies}; this extension requires the use of  ``good'' substitutes of the $L^p$ spaces for $p <1$ which are supplied by  the real Hardy spaces $H^p$. The important practical result   is that the $p$-leaders associated with this notion also are given by \eqref{pleaders}.

In applications,  one first computes the exponent $\Hmin $ and the function $\eta_f (p)$. If $\Hmin >0$, then one has the choice of using either $p$-leaders or wavelet leaders as multiresolution quantities. 
Though leaders are often preferred because of the simple interpretation that they yield in terms of  the most commonly used (H\"older) exponent, it has been remarked that $p$-leaders constitute a quantity which displays better statistical properties, because it is based on  averages of wavelet coefficients, instead of a supremum, i.e. a unique extremal value, see \cite{CRAS2019} and ref. therein. 
 If both $\Hmin <0$ and  $\eta_f (p) <0$ for all $p$s, then one cannot use directly these techniques and one performs a (fractional) integration on the data first. 
 If one wants to use wavelet leaders, the order of integration  $s$ has to satisfy $s > -\Hmin$ since $h^{min}_{f^{(-s)}}  = \Hmin +s$. 
 Similarly, in the case of  $p$-leaders it follows immediately from  the Sobolev interpretation \eqref{nicol2} of the wavelet scaling function
 that 
 \[ \eta_{f^{(-s)}}(p)   = ps + \eta_f(p). \] Thus, if  $\eta_f(p) <0$, then an analysis based on $p$-leaders will be valid if the order of fractional integration  $s$ applied to $f$ satisfies  $s > -\eta_f(p)/p$.  
 In practice, one does not perform a fractional integration on the data, but one simply replaces  the wavelet coefficients $c_{j,k}$ by 
 $2^{-sj}c_{j,k}$, which leads to the same scaling functions \cite{MandMemor}, and has the advantage of being performed at no extra computational cost.

 \subsection{Towards a classification of pointwise singularities} 
 
 \label{classif}
 

 In Section \ref{Poitexp} we motivated the introduction of alternative pointwise regularity exponents by the requirement of having a tool  available  for non locally bounded functions, which allows to deal directly with the data without having recourse to a smoothing procedure first;  but this variety of exponents can also serve another purpose: By comparing them, one  can draw differences between several types of singularities. This  answers an important challenge in several areas of science; for example, in fully developed turbulence, some models predict the existence of extremely oscillating structures such as \eqref{chirp} and  the  key signal processing problem for the detection of gravitational waves  also involves the detection of pointwise singularities similar to \eqref{chirp}  in extremely noisy data \cite{Fland}.    
 
  Let us start with a simple example:  Among the functions which satisfy $h_f (x_0) = \al $, the most simple pointwise singularities are supplied by { \em cusps}  singularities, i.e. by functions which ``behave'' like  
\BE \label{cusp}   \mathcal{C}_{\alpha}(x)=  | x-x_0|^\al  \qquad \mbox{ (if  $ \al >0$ and $\al \notin 2 \NN$)} .  \EE 
How can we ``model'' such a behavior?  A simple answer consists in  remarking that the primitive of \eqref{cusp} is of the same form, and so on if we iterate integrations. Since the mapping $t \rightarrow  h_{f^{(-t)}}(0)$ is concave \cite{ABJM}, it follows that \eqref{cusp} satisfies 
\[ \forall t >0, \qquad h_{\mathcal{C}^{(-t)}_{\alpha}} (t_0) = \al +t . \]
For cusp singularities, the pointwise H\"older exponent is exactly shifted by the order of integration. This is in sharp contrast with  the chirps   \eqref{chirp}, for which a simple integration by parts yields  that the H\"older exponent of its $n$-th iterated primitive is  
\[ \forall n \in \NN , \qquad h_{\mathcal{C}^{(-n)} _{\alpha, \beta }} (t_0) = \al + (1 + \be ) n  , \]
from which it easily follows that the fractional primitives of the chirp satisfy 
\[ \forall t >0, \qquad h_{\mathcal{C}_{\alpha, \beta}^{(-t)}} (t_0) = \al + (1 + \be ) t , \]
 \cite{ABJM}. 
We conclude from these two typical examples that inspecting simultaneously the  H\"older exponents of $f$ and its primitives, or its fractional integrals,   allows to put in light that oscillating behaviour of $f$  in the neighbourhood of its singularities which is typical of \eqref{chirp} (see \cite{LVS} for an in-depth  study of the information revealed by the mapping $t \rightarrow h_{f^{(-t)}} (t_0)$). 
To that end, the following definition was proposed, which encapsulates  the relevant ``oscillatory'' information contained in this function, using a single parameter.  

\BD\label{def.2} Let $f: \R^d \rightarrow \R $   be  such that $f \in L^p_{loc} $. If $h^p_f(x_0) \neq +\infty$, then the oscillation  exponent of 
$f$  at 
$x_0$  is   
\begin{equation} \label{eq:oscill}  \mathcal{O}_f (x_0)=    
\left( \frac{\partial}{\partial t} h^p_{f^{(-t)}} (x_0)  \right)_{t=0^+} -1 .
\end{equation} 
\ED

{ \bf Remark:} In theory,  a dependency in $p$ should appear in the notation since $f$  belongs to several $L^p$ spaces. However, in practice,  a given $p$ is fixed, and this inaccuracy does not pose problems. 
 
The choice of taking the derivative at $t=0^+$ is motivated by a robustness argument: The exponent should not  be perturbed when adding to $f$  a smoother term, i.e.  a term that would be a $O( |x-x_0|^{h})$ for an $h > h_f(x_0) $; it is a consequence of the following lemma, which we state in the setting of H\"older exponents (i.e. we take $p = + \infty$ in Definition \ref{def.2}).

\BL  Let $f$ be such that $h_f (x_0)  < +\infty$ and  $\mathcal{O}_f (x_0) < +\infty $; let  $g\in C^\al (x_0) $  for an $\al > h_f (x_0)$. Then, for  $s$ small enough,  the H\"older exponents of $(f+g)^{(-s)} $ and of $ f^{(-s}) $ coincide.
\EL

{ \bf Proof:} By the concavity of the mapping  $s \rightarrow h_{f^{(-s)}} (x_0)$, see \cite{Bandt2015,Porqu2017}, it follows that 
\[  h_{f^{(-s)}} (x_0)  \leq  h_{f} (x_0) + (1+ \mathcal{O}_f (x_0))s; \]
but one also has $ h_{g^{(-s)}} (x_0)  \leq  \al +s$; so that, for $s$ small enough, 
$ h_{g^{(-s)}} (x_0)   >  h_{f^{(-s)}} (x_0) $, and it follows that $h_{(f+g)^{(-s)}} (x_0)  = h_{f^{(-s)}} (x_0) $. 
\vspace{0.5cm} 

The   oscillation  exponent takes the value $\beta$ for a chirp; it is the first of { \em second generation exponents } that  do not measure a regularity, but yield additional information, paving the way to a   
richer description of  singularities.  In order to go further in this direction, 
we consider another  example:  { \em Lacunary combs}, which were first considered in   \cite{Bandt2015,Porqu2017}  (we actually deal here with a slight variant). 
Let $\phi =\mathbbm{1}_{[0, 1]} $.   
\BD 
Let $\alpha \in \RR  $ and $\gamma  >\ome >0 $. The { lacunary comb  }   $F^\al_{ \ome, \gamma}$, is 
\BE \label{defaog}  F^\al_{ \ome, \gamma} (x) = \sum_{j=1}^\infty 2^{-\al j} \phi \left( 2^{\gamma j} (x-2^{-\ome j})  \right) .  \EE
\ED

We consider its behaviour near the singularity at $x_0 =0$: 
    if $\al > -\gamma$, then $F^\al_{ \ome, \gamma} \in L^1 (\RR) $  and it is locally bounded if and only if $\al \geq 0$. In that case,  
one easily  checks  that 
\begin{equation} h_{F^\al_{ \ome, \gamma}}(0) = \frac{\al}{\ome}, \quad \mbox{and} \quad h_{{F^\al_{ \ome, \gamma}}^{(-1)}}(0) =\frac{\al + \gamma}{\ome} \end{equation} 
and one obtains (see \cite{Porqu2017}) that 
$\mathcal{O}_{F^\al_{ \ome, \gamma}}(0) =\frac{ \gamma}{\ome} -1 .  $

 We conclude that chirps and lacunary combs are two examples of oscillating singularities. They are,
 however, of  different nature: In the second case, oscillation is due to the fact that this
 function vanishes on larger and larger proportions of small balls centered at the origin (this is
  detailed in \cite{Bandt2015}, where this phenomenon is precisely quantified through the
 use of { \em accessibility  exponent } of a set at a point).  
On the other hand, chirps are oscillating singularities for a  different reason:  It is due to very fast oscillations, and compensations of signs. This  can be checked by verifying that   the oscillation exponent of  $| \mathcal{C}_{\alpha,\beta}| $  at $0$ vanishes.

  We will now see that  this difference can be put in evidence by considering the variations of the $p$-exponent. 
  Comparing the $p$-exponents of chirps and lacunary combs  allows  to draw a distinction between  their singularities; indeed,  for $p\geq 1$, see \cite{PART1}, 
\BE\label{eq:pFalphagamma}
h_{F^\al_{ \ome, \gamma}}^{p}(0)=\alpha+\frac{1}{p}\left(\frac{\gamma}{\ome}-1\right)
\EE
 whereas a straightforward computation yields  that 
\[ \forall p , \qquad h^{p}_{\mathcal{C}_{\alpha,\beta}}(0)=\alpha . \] 
We conclude that the $p$-exponent of $F^\al_{ \ome, \gamma}$  varies with $p$, whereas the one of $\mathcal{C}_{\alpha,\beta}$ does not.
We will introduce another pointwise exponent which captures the lacunarity of the combs; it requires first the following notion: 
If $f \in L^p_{loc}$ in a neighborhood of $x_0$ for $p >1$, the  {   \em critical Lebesgue index} of $f$ at $x_0 $ is
 \BE  \label{critleb} \pfx   = \sup \{ p: f \in  L^{p}_{loc}  (\RR)  \mbox{ in a neighborhood of   $x_0$}  \} .  \EE  
 The $p$-exponent at $x_0$ is  defined on the  interval  $[1, \pfx  ]$ or $[1, \pfx  )$.  We denote: 
 $ \qfx ={1}/{\pfx}.$
Note that $\pfx$ can take the value $+\infty$. An additional pointwise exponent, which, in the case of lacunary combs, quantifies the sparsity of the ``teeth''  of the comb, can be defined as follows see \cite{Bandt2015}. Its advantage is that it quantifies the ``lacunarity information'' using a single parameter instead of the whole function $p\rightarrow {h}^{ (p)}_f (x_0)$.

\BD  \label{deflac} Let  $f \in  L^p_{loc}$ in a neighborhood of $x_0$ for a $p >1$. The {  lacunarity exponent } of $f$ at $x_0$  is
 \BE  \label{eq:lac_exp} { L}_f (x_0) =
  \frac{\partial }{\partial q} \left(  {h}^{ (1/q)}_f (x_0) \right)_{ q={\qfx}^+}. \EE 
 \ED
 
 This quantity  may have to be understood as a limit when $q \to \qfx$, since $h_f^{1/q} (x_0)$ is not necessarily  defined for $q=\qfx $.  This limit always exists as a consequence of the concavity of the mapping $q \rightarrow h_f^{1/q} (x_0)$, and it is nonnegative (because this mapping is increasing).

 The lacunarity exponent of  $F^\al_{ \ome, \gamma}$ at $0$ is $\frac{\gamma}{\ome}-1$,  which
 puts into light the fact  that this exponent allows to measure  how  $F^\al_{ \ome, \gamma}$
 vanishes on  "large sets" in the neighborhood of $0$ (see \cite{Bandt2015} for a precise
 statement). Furthermore the {  oscillation  exponent } of   $F^\al_{ \ome, \gamma}$ at $0$ is  
 $\frac{\gamma}{\ome} - 1$, so that it coincides with the lacunarity exponent. 
 The  {  oscillation  exponent }    is always larger than the lacunarity exponent. A way to distinguish between the effect due to lacunarity and the one due to cancellations is to introduce a third exponent, the { \em cancellation exponent}  \[ {\mathcal C}_f (x_0)=  {\mathcal O}_f (x_0) - {L}_f (x_0).   \]  
 The  lacunarity and the cancellation exponents   lead to the following classification  of pointwise singularities see \cite{Porqu2017}.

\BD
Let $f$ be a tempered  distribution on $\RR$:   \begin{itemize}
\item $f$ has a { \bf canonical singularity}  at $x_0$ if ${\cal O}_f(x_0)= 0 $.
\item $f$ has a { \bf balanced singularity}  at $x_0$ if ${L}_f(x_0)= 0 $  and $\mathcal{C}_f(x_0)\neq  0 . $
\item  $f$ has a { \bf lacunary singularity} at $x_0$  if $\mathcal{C}_f(x_0)= 0  $ and ${L}_f(x_0) \neq  0 $.   
\end{itemize}
\ED
Cusps are  typical examples of    canonical singularities, 
chirps are  typical examples of  balanced singularities and 
lacunary combs are    typical examples of  lacunary singularities. 
 
 Many probabilistic models display lacunary singularities: It is the case e.g. for random wavelet series \cite{Bandt2015,Porqu2017}, some L\'evy processes, see \cite{Balanca} or fractal sums of pulses \cite{Saes}. Note that our comprehension of this phenomenon is  very partial: For instance, in the case of L\'evy processes, the  precise  determination of the conditions that a  L\'evy measure should satisfy in order to  guarantee  the existence of lacunary singularities has not been worked out: in \cite{Balanca}, P. Balanca proved that some self-similar L\'evy processes with even L\'evy measure  display oscillating singularities, which actually turn out to be lacunary singularities  and also that L\'evy processes which  have only positive jumps do not display such singularities;  and, even in these cases, only a lower bound on  their Hausdorff dimensions has been obtained.  In other words, for L\'evy processes,  a joint multifractal analysis of the H\"older and the lacunarity exponent remains to be worked out.  Note also that there exists much less examples of functions with  balanced singularities:  In a deterministic setting it is the case for the Riemam function \cite{JaffMey1}  at certain rational points. However, to our knowledge, stochastic processes with balanced singularities have not been met up to now.

Another important question is to find numerically  robust ways to determine if a signal has points where it displays balanced or   lacunary singularities. This question is important in several areas of physics; for instance,  in hydrodynamic turbulence, proving   the presence of oscillating singularities would  validate certain vortex stretching mechanisms which have been proposed, see  \cite{Frisch1995}.  Another motivation  is methodological: if a signal only has  canonical singularities,  then its $p$-multifractal spectrum does not depend on $p$  and  its singularity spectrum is translated by $t$ after a fractional integral of order, so that all methods that can be used to estimate its multifractal spectrum yield the same result (up to a known shift in the case of a fractional integration).  An important questions related with the multifractal formalism is to determine if some of its variants allow to throw some light on these problems. Motivated  by applications to  physiological data, we shall come back to this question in Sections \ref{Marathon1} and  \ref{intfrac}. 

Note that the choice of three exponents to characterize the ``behaviour'' of a function in the neighbourhhood of one of its singularities may seem arbitrary; indeed, one could use the very complete information supplied by the following  two variables function:
 If $f$ is a tempered distribution, then the  {\em  fractional exponent} of $f$ at $x_0$  is the  two variable function
\[ { \mathcal H}_{f, x_0} (q,t ) =  h^{1/q}_{f^{(-t)}} (x_0) -t, \]  see \cite{Porqu2017} where this notion is introduced and its properties are investigated. However, storing the pointwise regularity behaviour through the use of a two-variables function defined at every point is unrealistic, hence the choice to store only the information supplied by the three parameters we described.  
 This choice is   motivated by two conflicting requirements:  On one hand, one wishes  to introduce mathematical tools  which are sophisticated enough to describe several ``natural'' behaviours that can show up in the data, such as those supplied by  cusps, chirps, and lacunary combs. On other hand,  at the end,  classification has to  bear on as little parameters as possible in order to be of practical use in applications; the goal here is to introduce a multivariate multifractal analysis based on a single function $f$, but applied to  several pointwise exponents associated with $f$ (say two or three among a  regularity, a lacunarity and a cancellation exponent).

 Our theoretical comprehension of   which  functions can be  pointwise exponents   is extremly partial, see  \cite{SeuretPres} for a survey on this topic: It has been known for a long time that a pointwise H\"older exponent $h_f (x)$ can be any nonnegative function of $x$ which can be written as a liminf of a sequence of continuous functions, see \cite{Jaffard1995,AyaJaff,Daoud}, but the same question for $p$-exponents is  open (at least in the case where it takes negative values). Similarly, which couples of functions $(h(x), O(x))$ can be the joint H\"older and oscillation exponents of a function also is an open question (see \cite{JaffReV} for partial results), and it is the same  if we just consider the oscillation exponent, or  couples including the lacunarity exponent.   One meets similar limitations for multifractal spectra:  In the univariate setting supplied by the multifractal H\"older spectrum, the general form of functions which can be multifractal spectra is still open; nonetheless  a partial result is  available: functions which can be written as infima of a sequence of continuous functions are multifractal spectra \cite{JaffFrac}; additionally, as soon as two exponents are involved, extremly few results are available.  For instance, if $f$ is  a locally bounded function, define its { \em  bivariate oscillation spectrum}  as 
\[ { \cal D}_f (H, \beta ) = \dim \{ h: h_f (x) = H \quad \mbox{and}  \quad {\cal O}_f (x) = \beta \} . \] 
Which  functions of two variables $D (H, \beta)$ can be bivariate oscillation spectra is a completely open problem.

 \subsection{Mathematical results concerning the multifractal formalism} 
 
 \label{mathresmf}
 
 We now  consider a  general setting where  $h: \RR \rightarrow \RR$ is a pointwise  exponent   derived   from  a multiresolution quantity $ d_{\lambda} (= d_{j,k} )$  according to Def. \ref{derived}, and defined in space dimension $d$.
  The associated  multifractal spectrum  
 ${\cal D} $ is 
 \[  {\cal D}(H) = \dim(  \{ x: \quad h(x) = H\} ). \] 
The { \em support} of the spectrum is  the image of the mapping $x\rightarrow h(x)$, i.e. the collection of values   of $H$ such that \[ \{ x \in \RR:  h(x) = H \} \neq \emptyset  \]
(note that this denomination, though commonly used, is misleading, since it may not coincide with the mathematical notion of support of a function). 

 The    { \em  leader scaling  function} associated with the multiresolution quantities  $(d_{j,k})$  is 
 \BE \label{defscalond2}  \forall q \in \RR, \hspace{6mm} 
 \zeta_f (q) =   \displaystyle \liminf_{j \rightarrow + \infty} \;\; \frac{\log \left( 2^{-j} \displaystyle\sum_{k }  | d_{j,k}  |^q.  \right) }{\log (2^{-j})}. \EE 
 Note that, in contradistinction with the wavelet scaling function, it is also defined for $p <0$.  Referring  to ``leaders'' in the name of the scaling function does not mean  that the $d_{j,k}$ are necessarily obtained as wavelet leaders or wavelet $p$-leaders, but  only to prevent any  confusion with the wavelet scaling function. 
The  { \em Legendre spectrum}  is 
\BE \label{lspec} {\cal L} (H)  : =  \inf_{q \in \RR} (1+qH - \zeta_f(q)) .  \EE 
As soon  as  relationships such as   (\ref{carachqf})  hold, then the following upper bound is valid
\begin{equation}
\label{equ-tl} \forall H, \qquad
{\cal D}(H) \leq {\cal L} (H)   
\end{equation}
(see \cite{Jaffard2004} for particular occurrences of this statement, and \cite{Abel} for the general setting).  However,  for  a number of synthetic processes with known  ${\cal D}(H) $ (and for a proper choice of the multiresolution quantity), this  inequality turns out to be an equality, in which case, we will say that the { \em multifractal formalism holds}. The leader scaling functions obtained using wavelet leaders or $p$-leaders can be shown to enjoy the same robustness properties as listed at the end of Section  \ref{Univariate3}, see \cite{Abel} (it  is therefore also the case for the  Legendre spectrum).    It follows from their mathematical  and numerical properties  that wavelet leader based techniques   form the state of the art  for real-life signals multifractal analysis. 

\begin{figure}
    \centering
    \includegraphics[scale=0.3]{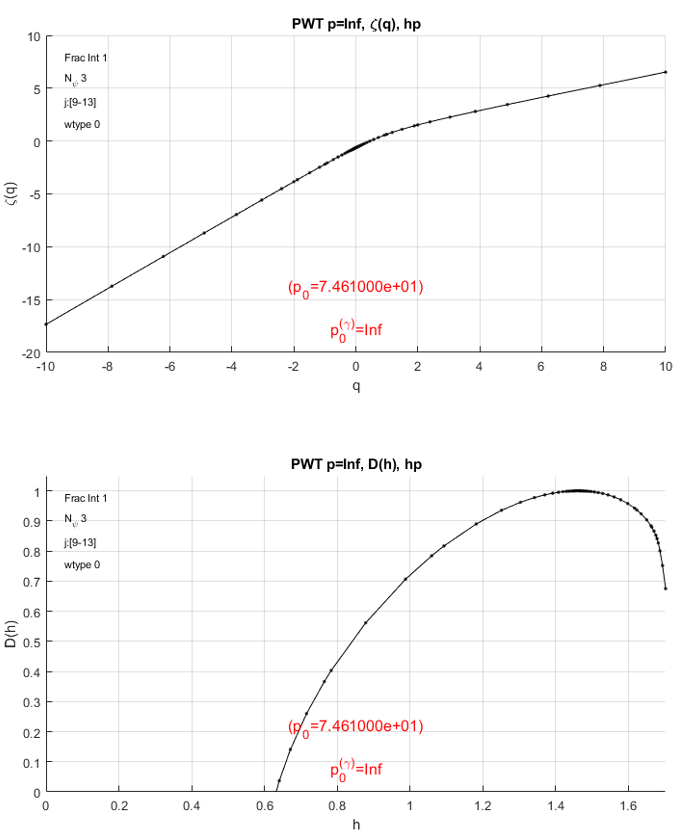} \qquad
    \includegraphics[scale=0.3]{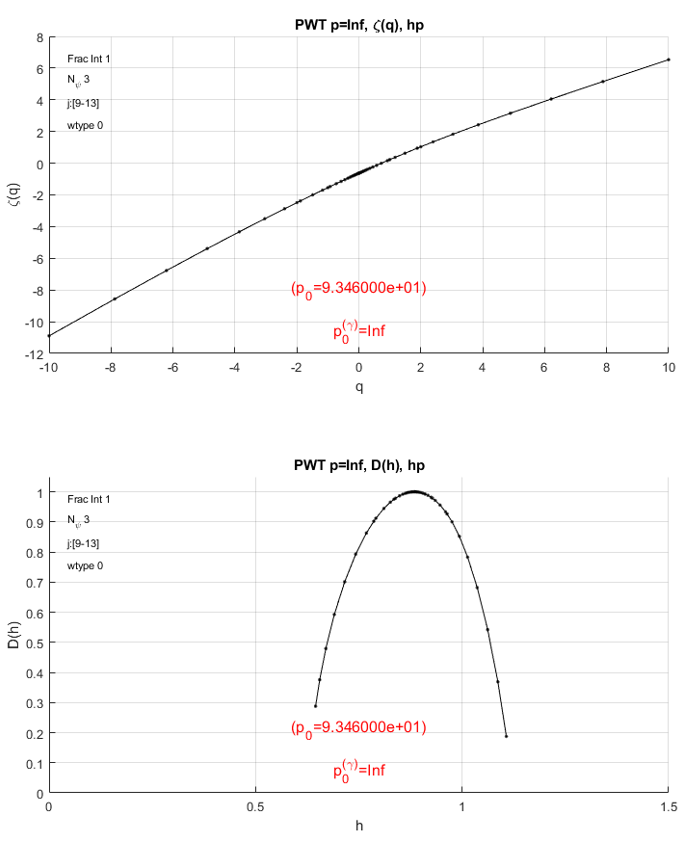}
    \caption{Representation of scale function and the univariate H\"older Legendre spectra of the primitives of heart beat frequency (left) and cadence (right)  of one marathon runner during the entire race. the multiresolution quantities used in these derivation are the wavelet leaders of the primitive of the data}   
    \label{bilan_pl3}
\end{figure}

  In applications,   one cannot  have access to the regularity exponent at every point in a numerically stable way, and thus  ${\cal D}(H) $ is unaccessible; this explains why, in practice,   ${\cal L} (H) $  is the only computationally available spectrum,  and  it is  used as such in  applications. 
   However,   information on the pointwise exponent may  be inferred from the Legendre spectrum. Such results are collected in the following theorem, where  they  are stated in decreasing order of generality.
   
   \BT  \label{theo1} Let $h: \RR \rightarrow \RR)$ be a pointwise  exponent, and assume that it is  derived   from multiresolution quantities $d_{j,k}$  according to Def. \ref{derived}.  The following results on $h$ hold:
 \begin{itemize}
 \item Let 
 \BE 
\label{caracbeswav3hol2}  
h^{min}   =\liminf_{j \rightarrow + \infty} \;\;  
 \;\;  \frac{ \log 
\left(  \displaystyle  \sup_{  k}  d_{j,k}   \right) }{\log (2^{-j})} \quad \mbox{ and } \quad h^{max}  =\liminf_{j \rightarrow + \infty} \;\;  
 \;\;  \quad  \frac{ \log 
\left(  \displaystyle  \inf_{  k}  d_{j,k}  \right) }{\log (2^{-j})}
\EE
then 
 \BE 
\label{encadrhol}  \forall x \in \RR \qquad h^{min} \leq h(x) \leq h^{max}  . \EE
\item If the Legendre spectrum has a unique maximum  for  $H= c_1$, then 
\BE \label{almoster} \mbox{ for almost every } x, \qquad h (x) = c_1;\EE 
\item If the leader scaling function \eqref{defscalond2} associated with the $d_{j,k}$ is affine, then  $f$ is a monoh\"older function, i.e. 
  \[ \exists H_0: \quad \forall x, \qquad h (x) =H_0,\] 
  where $H_0$ is the slope of the leader scaling function.
\end{itemize}
\ET 

{\bf Remark:}  The last statement asserts that, if $h$ is a pointwise exponent associated with a function $f$, then $f$ is a monoh\"older function. This result has important implications in modeling since it yields a numerically simple test, based on global quantities  associated with the signal, and which yields the pointwise exponent everywhere. This is in strong contradistinction with the standard pointwise regularity estimators, see e.g. \cite{Bardet} and ref. therein, which are based on local estimates, and therefore on few data thus showing strong statistical variabilities, and additionally often assume that the data follow some a priori models.
\vspace{0.5cm}  

{\bf Proof:} We first prove the upper bound in (\ref{encadrhol}). Let $\al > h^{max}$; there exists a sequence $j_n \rightarrow + \infty $ such that 
\[ \log \left(\inf_{ k }  d_{j_n,k} \right) \geq \log (2^{-\al j_n}), \] 
so that at the scales $j_n$ all $d_\la$ are larger than $2^{-\al j_n}$. It follows  from (\ref{carachqf}) that 
\[ \forall x, \qquad h (x) \leq \al , \] and the upper bound follows. 
The proof of the  lower bound is similar (see e. g. \cite{Zuhai}).  

The second statement  is direct consequence of the following upper bounds for the dimensions  of the sets 
\BE  \label{uppersets} E_{H}^+ = \{ h (x) \geq H \} \quad \mbox{ and } \quad E_{H}^- = \{ h (x) \leq H \}  \EE
 which are a  slight improvement of (\ref{equ-tl}), see \cite{Zuhai}: 
   
   \BP Let $h$ be a pointwise exponent derived from the multiresolution quantity $( d_{j,k})$. Then the following bounds hold:
  \BE \label{majdim3}\dim (E_{H}^-)   \leq   \inf_{q >0} (1+qH - \zeta_f(q)) \quad \mbox{ and }  \quad \dim (E_{H}^+) \leq  \inf_{q<0} (1+qH - \zeta_f(q)) \EE
   \EP
Let us check how \eqref{almoster} follows from this result. 
Note that the first (partial) Legendre transform yields the increasing part of ${\cal L}(H)$ for $H \leq c_1$  and the second one yields the decreasing part  for $H \geq c_1$.  If ${\cal L}$ has a unique maximum  for  $H= c_1$,  it follows from (\ref{majdim3}) that
\[ \forall n , \qquad \dim (E_{c_1-1/n}^-) <1 \quad \mbox{ and  } \quad  
\dim (E_{c_1+1/n}^-) <1.  \]
All of these sets therefore have a vanishing Lebesgue measure, which is also the case of their union. 
But this union is  $\{ x: h(x) \neq c_1\} $. It follows that almost every $x$ satisfies $h(x) = c_1$. 

Finally, if  the leader scaling function is affine, then  its Legendre transform is supported by a point $H_0$ and takes the value $-\infty$ elsewhere. The upper bound (\ref{equ-tl}) implies that, if $H \neq H_0$ the corresponding isoregularity set is empty. In other words, $H_0$ is the only value taken by the pontwise exponent, and   $f$ is a monoh\"older function.
\vspace{0.5cm} 

{ \bf Remarks:} 

If $h^{min}= h^{max}$, the conclusion of the first and last statement are the same.  However, one can check that the condition $h^{min}= h^{max}$ is slightly less restrictive than requiring the leader scaling function to be  affine (the two conditions are equivalent if, additionally,  the $\liminf$ in (\ref{caracbeswav3hol2})  is  a limit).

   The parameter $c_1$ defined in Theorem \ref{theo1} can be directly estimated using log-log plot (see \cite{MandMemor} and ref. therein), and, in practice  it plays an important role in classification as we will see in the next section. When the multiresolution quantity used is  the $p$-leaders of a function $f$, the associated exponent $c_1$ may depend on $p$, and we will mention this dependency and denote this parameter by $c_1(p, f)$.   This is in contradistinction with the exponent $h^{min}$  defined by (\ref{caracbeswav3hol2}), which, in the case of functions with some uniform H\"older regularity,   coincides  with  the exponent $\Hmin$ defined by (\ref{caracbeswav3hol}) for leaders and $p$-leaders, as shown by the following lemma; note that it is actually preferable to   compute it   using  (\ref{caracbeswav3hol}), which has the advantages of being well defined without any a priori assumption on $f$.

   \BL Let $f: \RR \rightarrow \RR  $ be such that  $\Hmin >0$. Then  the $h^{min} $ parameter computed using $p$-leaders all coincide with  the $\Hmin $ computed using wavelet coefficients. 
   \EL

   Let us sketch the poor of this result.  Suppose that $\Hmin >0$ and let  $\al>0$ be such that  $\al < \Hmin$. Then, the wavelet coefficients of $f$ satisfy 
   \[ \exists C, \quad \forall j,k \qquad | c_{j,k} | \leq C 2^{-\al j} . \]
   Therefore the $p$-leaders of $f$ satisfy 
   \[  
  \pL_\la \leq  \left(   \sum_{ \lambda' \subset 3 \lambda} (2^{-\al j'})^p \, \, 2^{j-j'} \right)^{1/p}  \]
  \[ \leq  \left(   \sum_{ j' \geq j } 2^{-\al pj'} \, \, 2^{j-j'} \right)^{1/p}  \leq C 2^{-\al j} ;\]   it follows that   the corresponding $p$-leader is smaller that $|c_{\la_n}|$  so that  the $h^{min} $ computed using $p$-leaders is  smaller that the one computed using wavelet coefficients. 
   Conversely,  by definition of $\Hmin $, there exists a sequence of dyadic intervals  $c_{\la_n}$ of width decreasing to 0, and such that 
   \[ |c_{\la_n}| \sim 2^{-\Hmin j_n} , \] 
  and the corresponding $p$-leader is larger that $|c_{\la_n}|$  so that  the $h^{min} $ computed using $p$-leaders is  smaller that the one computed using wavelet coefficients.  
  \vspace{0.5cm}
   
The following result yields an important a priori bound on the dimensions of the singularity sets corresponding to negative regularity exponents, see \cite{PART1}. 

\BP 
Let $p > 0$,   and let  $f: \RR \rightarrow \RR$ be a function such that $\eta_f (p) >0$. Then its $p$-spectrum satisfies
\[ \forall h , \qquad  { \mathcal D}_p (H)  \leq 1 + Hp\] 
\EP

   Let us elaborate  on the information supplied by the exponent $c_1 (p,f)$:   A direct consequence  of \eqref{almoster} is that, if  a signal $f$ satisfies
 that the exponent $c_1 (p,f)$  takes the same value for  $p_1 <  p_2$,  then  this implies that the $p$-exponent satisfies that 
  \[  \mbox{ for almost every } x,  \qquad h^{p_1}_f (x) =  h^{p_2}_f (x) ,\]  
  which implies that  the mapping $p \rightarrow h^{p_1}_f (x)$ is constant for $p \in [p_1, p_2]$; but, since the mapping $p\rightarrow h^{1/p}_f (x_0) $ is concave and increasing,  see \cite{Bandt2015,Porqu2017},  it follows that this mapping is constant  for $p$ small enough; 
    as a consequence, the lacunarity exponent vanishes at $x$. 
  Similarly, if, for a given $p$,   $c_1 (p,f^{(-1)}) -c_1 (p,f) =1$, this implies that 
   \[  \mbox{ for almost every } x,  \qquad h^p_{f^{(-1)}} (x) =  h^p_{f} (x) +1,\] 
   and the same argument as above, see \cite{Bandt2015,Porqu2017}, yields  the absence of oscillating singularities for almost every point.    
   In other words, the computation of $c_1(p)$ yields a key information on the nature of the singularities a.e. of the signal, which we sumarize in the following statement, which will have implications in the next section for the analysis of  marathon runners data. 
  
  \BP \label{propae} Let  $f: \RR \rightarrow \RR$  be  a function in $L^p$.  
  
   If 
  \[ \exists  q > p  :  \qquad c_1 (p, f) = c_1 (q, f),\]
  then  for almost every $x$,  $f$ has no lacunary singularity at $x$.
  
  If  $f$ satisfies
  \[ \exists p    : \qquad c_1 (p,f^{(-1)}) -c_1 (p,f) =1, \]  
  then,  for almost every $x$,  $f$ has a canonical  singularity at $x$.  
  \EP

 These two results are characteristic  of signals  that only contain canonical  singularities,
 see Section \ref{classif}, and they also demonstrate that $c_1(p, f)$, which, in general,   depends on the value of $p$ is intrinsic for such data (see a contrario  \cite{Bandt2015} where the exponent $c_1(p, f)$ of lacunary wavelet series is shown to depend on the value of $p$, and \cite{Saes} where the same result is shown for random sums of pulses).   Note that such results  are available in the discrete wavelet approach only; they would not be possible using the WTMM or the MFDFA approaches, which do not allow to draw differences between various pointwise regularity exponents and therefore  do not yield spectra fitted to different values of the $p$-exponent. 
 To summarize, the  advantages of  the $p$-leader based multifractal analysis framework   are: the  capability to estimate negative regularity exponents, better estimation performances, and a  refined characterization of the nature of pointwise regularities.
 
  One important argument in favor of multifractal analysis is that it supplies robust classification parameters, in contradistinction with   pointwise regularity which can be extremely erratic.  Consider for instance the example of a sample path of a L\'evy process without Brownian component (we choose this example because such processes now play a key role in statistical modeling):   Its H\"older exponent  is a random,   everywhere discontinuous, function which cannot be numerically estimated or even drawn \cite{JaffLev}: In any arbitrary small interval $[ a, b]$ it takes all possible values $H\in [0,  H^{max}] $. On the opposite, the multifractal spectrum (which coincides with the Legendre spectrum) is extremly simple and robust to estimate numerically: It is a deterministic linear  function on the interval $[0, H^{max}] $ (with $D(H^{max}) = 1$).  This example is by no means accidental: though one can simply construct stochastic processes with a random multifractal spectrum (consider for instance a Poisson process restricted to an interval of finite length),  large classes of classical processes have simple deterministic multifractal spectra (and Legendre spectra), though no simple assumption which would guarantee this results is known. The determination of a kind of ``0-1 law'' for multifractal spectra, which would guarantee that, under fairly general assumptions,  the spectrum almost surely is  a deterministic function, is a completely open problem, and  its resolution would greatly improve our understanding of the subject. Even in the case of Gaussian processes, though it is known that such processes can have a random H\"older exponent \cite{AyaGauss},  the possibility of having a random multifractal spectrum  still is   a open issue. 
 
 \subsection{Generic results} 

\label{gener}

Let us come back to the problem raised in Section \ref{Univariate1}  of estimating the size of the H\"older singularity sets of increasing functions which  led us to the key  idea that the Hausdorff dimension is the natural way to estimate this size. One can wonder if the estimate \eqref{majspecinc} that we found for the multifractal spectrum is optimal. In 1999, Z. Buczolich  and J. Nagy answered this question in a very strong way, showing that it is sharp for a { \em residual} set of continuous increasing functions, see \cite{BucNag}. What does this statement precisely mean?  Let  $E$ be the set of continuous increasing functions $f: \;  \RR \rightarrow \RR$, endowed with the natural distance supplied by the $\sup$ norm. Then  equality in \eqref{majspecinc} holds  (at least)  on a residual set in the sense of Baire categories, i.e. on a countable intersection of open dense sets. 

This first breakthrough opened the way to genericity results in multifractal analysis.  They were the consequence of the important remark that  scaling functions for $p >0$ can be interpreted as stating  that  $f$ belongs to an intersection of Sobolev spaces $E_\eta$ (in the case of the Kolmogorov scaling function) or of a  variant of these spaces, the { \em oscillation spaces} in the case of the leader scaling function \cite{Jaf05}.  One easily checks that $E_\eta$  is a complete metric space,  and   the Baire property therefore is valid (i.e. a countable intersection of open dense sets is dense).  The question formulated by Parisi and Frisch in \cite{ParFri85},
can be reformulated in this setting:  
If equality in  \eqref{equ-tl} cannot hold for { \em every}  function in $E_\eta$ (since e.g. because it contains $C^\infty$  functions), nonetheless it  holds on a residual set
\cite{JafBaire}. This result found many extensions: The first one consists in replacing the  genericity notion supplied by  Baire's theorem by the more natural notion supplied by { \em prevalence}, which is an extension, in infinite dimensional function spaces of the notion of ``Lebesgue almost everywhere'', see \cite{Christ,HSY} for the defintion of this notion and its main properties, and \cite{FJ} for its use in the setting of multifractal analysis.  The conclusions drawn in the Baire setting also hold in the prevalence setting, and raise the question of the determination of a stronger notion of genericity, which would imply both Baire and prevalence genericity, and which would be the ``right '' setting for the validity of the multifractal formalism. A natural candidate is supplied by the notion of { \em \hspace{-2mm} porosity }, see \cite{Linden},   but the very few results  concerning multifractal analysis in this setting   do not allow to answer this question yet. Note also that  Baire and  prevalence results have been extended to the $p$-exponent setting \cite{Fra}, which allows to deal with spaces of functions that are not locally bounded.   Another key problem concerning the generic validity of the multifractal formalism concerns the question of taking into account the information supplied by  negative values of $p$ in the scaling function \eqref{defscalond2}.  The main difficulty here is that the scaling function does not define a function space any longer, and    the  ``right''    notion of genericity which should be picked is competely open: Though Baire and prevalence do not really require the setting supplied by a (linear) function space, nonetheless these notions are not fitted to the setting supplied by a given scaling function which includes negative values of $p$. In \cite{BaSeurGen} J. Barral and S. Seuret developed an alternative point of view which is less ``data driven'': They reinterpreted the question in the following way:     Given a certain scaling function $\eta (p)$, they considered the problem of  constructing an ad hoc  function space which is taylored so that  generically (for the Baire setting), functions in such a space  satisfy the multifractal formalism for the corresponding scaling function, including its values for $p <0$ (and Legendre spectrum).  Another limitation of the mathematical results of genericity at hand is that they are not able to take into account { \em selfsimilarity} information: In \eqref{defscalond}, in order to  introduce a quantity which is always well-defined, and corresponds to  a function space regularity index,  the scaling function is defined by a $\liminf$. But,  most of the time, what is actually observed on the data (and what is really needed in order to obtain a numerically robust estimate) is that this $\liminf$ actually is  a true limit, which means that the $L^p$ averages of the  data display   exact power-law  behaviours at small scales. Up to now, one has not been able to incorporate  this type of  information  in the function space modeling developed.

\subsection{Implications on the analysis of marathon runners data} 

\label{Marathon1}


The increasing popularity of marathons today among all ages and levels is inherited from the human capacity to run long distances using the aerobic metabolism
\cite{lieberman07a}, which led to a  
 rising number of amateur marathon runners  who end the 42,195 km  between  2h40min and 4h40min. Therefore, even if nowadays, marathon running becomes “commonplace”, compared with ultra-distance races, this mythic Olympic race is considered to be the acme of duration and intensity \cite{Maron}. Running a marathon remains scary and complex due to the famous  “hitting the wall” phenomenon, which is the most iconic feature of the marathon \cite{Berndsen}. This phenomenon was previously evaluated in a large-scale data analysis of late-race pacing collapse in the marathon \cite{Smyth21};  \cite{Smyth18} presented an analysis of 1.7 million recreational runners, focusing on pacing at the start and end of the marathon, two particularly important race stages. They showed how starting or finishing too quickly could result in poorer finish-times, because fast starts tend to be very fast, leading to endurance problems later, while fast finishes suggest overly cautious pacing earlier in the race \cite{Smyth18}. 
Hence, the definition of a single marathon pace is based on the paradigm that a constant pace would be the ideal one. However, in \cite{Billat19},  a   3 years study shows that     large speed  and pace variations  are the best way to optimize performance. Marathon performance depends on pacing oscillations between non symmetric extreme values \cite{pycke2022marathon}. 
Heart rate (HR) monitoring, which reflects exercise intensity and environmental factors, is often used for running strategies in marathons. However, it is difficult to obtain appropriate	feedback for only the HR value since, as we saw above, the cardiovascular drift occurs during prolonged exercise. Therefore, now we have still to investigate whether this pace (speed) variation has a fractal behavior and if so, whether this is the case for the runners's heart rate which remains a pacer for the runners who aim to keep their heart rate in a submaximal zone (60-80 $\%$  of the maximal heart rate) \cite{Maron}.    
Here, we hypothesized that marathonians acceleration (speed variation),   cadence (number of steps per minute) and   heart rate time series follow a multifractal formalism and could be described by a self similar functions. 
Starting in the  1990s, many authors  demonstrated the fractal behavior of physiological data such as heart rate, arterial blood pressure, and breath frequency of human beings, see e.g.  \cite{ABRY:2010:A,ivanov1999}. In  2005, using the  Wavelet Transform Maxima Method,   E. Wesfreid, V. L. Billat and Y. Meyer \cite{wesfreid05a} performed  the first  multifractal analysis of  marathonians  heartbeats. This study was  complemented in 2009  using the   DFA (Detrended Fluctuation Analysis) 
 and wavelet leaders applied on a primitive of the signal \cite{wesfreid09a}.  
 Comparing the outputs of these analyses is hasardous; indeed, as already mentioned,  these methods are not  based on the same regularity exponents: WTMM is adapted to the  {\sl weak scaling exponent} \cite{MeyWVS},  DFA to the $p$-exponent for $p =2$  \cite{PART1,PART2},  and wavelet leaders to the  Hölder exponent  \cite{Jaffard2004}.
In the following, we will propose a method of digital multifractal analysis of signals based on $p$-leaders, which, in some cases, can avoid performing fractional integrations (or primitives) and thus transform the signal.
 In  \cite{wesfreid09a}, it was  put in evidence  that multifractal parameters associated with heart beat intervals evolve during the race  when the runner starts to be deprived of  glycogen (which is the major cause of the speed  diminution at the end of the race. 
This study also revealed that  fatigue decreases the running speed and
affects the regularity properties   of the signal  which can be related with the feelings of the runner measured by the Rate of Perception of Exhaustion (RPE),  according to the psychophysiological scale of Borg (mainly felt through the breathing frequency). 
In addition, there is a consistent decrease in the relationship between speed, step rate, cardiorespiratory responses (respiratory rate, heart rate, volume of oxygen consumed), and the level of Rate of Perception of Exhaustion (RPE), as measured by Borg's psychophysiological scale. The runner does not feel the drift of his heart rate, in contradistinction with his respiratory rate. These physiological data are not widely available and only heart rate and stride rate are the measures available to all runners for economic reasons. Moreover, these data are generated heartbeat by heartbeat and step by step. 

Our purpose in this section is to complement these studies by showing that a direct analysis on the data is possible if using  $p$-leaders  (previous studies using the WTMM or the standard leaders had to be applied to a primitive of the signal), and that they lead to a sharper analysis of the physiological modifications during the race. We  complement the previous  analyses  in order to demonstrate  the modifications of multifractal parameters during the race, and  put in evidence the physiological impact of the intense effort after the 20th Km. For that purpose, we  will perform   a  multifractal analysis  based on $p$-leaders.

We analyzed the heartbeat frequency  of 8 marathon runners (men in the same age area).
 Fig.\ref{hmin}  shows the determination of  exponents $\Hmin$ for  heartbeat frequency  and cadence  through a log-log regression; 
the regression is  always performed between the scales  j = 8 and  j = 11 (i.e. between  26s and  3mn 25s), which have been identified as the pertinent scales for such physiological data, see \cite{ABRY:2010:A}.
For most marathon runners,  $\Hmin$ is  negative,
see  Table \ref{tab1},  which justifies the use of $p$-leaders.   We then compute the wavelet scaling function in order to determine a common value of $p$ for which all runners satisfy 
 $\eta (p) >0$,   see Fig. \ref{scaling function} where examples of wavelet scaling function are supplied for  heartbeat frequency  and cadence. In the case of heartbeat frequency,   the computation of the 8 wavelet scaling functions yields that  $p = 1$ and $p= 1.4$  can be picked.    The corresponding $p$-leaders multifractal analysis is performed for these two values of $p$, leading to values of $c_1 (p)$ which are also collected in  Table \ref{tab1}.

 \begin{table}[htb]
    \caption{ Multifractal Analysis of heartbeat frequency of marathon runners  (Pr. : primitive)}\label{tab1}
    \begin{center}
    \resizebox{100mm}{!}{
    \begin{tabular}{||c||*{10}{m{2cm}|}|}
        \hline\hline
        $ $   & $H_{min}$  & $H_{min}$ of the Pr.  & $c_1$ for $p=1$ &  $c_1$ for $p=1.4$  & $c_1$ of the  Pr. for $p=1$ &   $c_1$ of the Pr. for $p=1.4$  \\ 
        \hline
        R1 & $-0,2768$ & $0,7232$ & $0,8099$ & $0,8064$ & $1,8242$ & $1,8213$  \\
        \hline
        R2 & $-0,0063$ & $0,9937$ & $0,4564$ & $0,4043$ & $1,3926$ & $1,3509$  \\
        \hline
        R3 & $-0,0039$ & $0,9961$ & $0,6856$ & $0,6625$ & $1,6942$ & $1,6351$ \\
        \hline
        R4 & $-0,1633$ & $0,8367$ & $0,6938$ & $0,6785$ & $1,6653$ & $1,6636$ \\
        \hline
        R5 & $-0,2434$ & $0,7566$ & $0,5835$ & $0,5689$ & $1,5401$ & $1,5224$ \\
        \hline
        R6 & $-0,3296$ & $0,6704$ & $0,5809$ & $0,5636$ & $1,5644$ & $1,5500$  \\
        \hline
        R7 & $0,1099$ & $1,1099$ & $0,5652$ & $0,5483$ & $1,4754$ & $1,4379$  \\
        \hline
        R8 & $-0,5380$ & $0,4620$ & $0,3382$ & $0,2977$ & $1,2588$ & $1,2086$  \\
        \hline\hline
    \end{tabular}}
    \end{center}
\end{table}

\begin{figure}
    \begin{center}
    \resizebox{80mm}{!}{
    \includegraphics{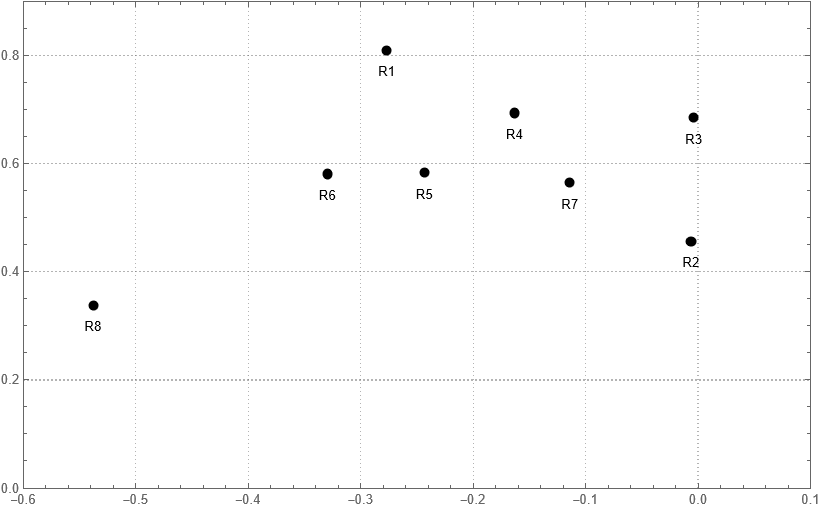}}
    \end{center}
    \caption{Representation of the pair $(H_{min}, c_1(p))$ with $p=1$ deduced from the 1-spectrum of heart rate and computed for the entire race; $H_{min}$ appears as the most relevant classification parameter. The isolated point on the left corresponds to R8, the most trained runner.}\label{Fig6}
\end{figure}

 In  Fig. \ref{Fig6}, the value of the couple  $(\Hmin, \cu )$ is plotted  (where we denote by $\cu$  the value of $H$ for which the maximum of the $p$-spectrum is reached). The values of $\cu  $ are very close to $0.4$  whereas the values of $\Hmin $  notably differ, and  are clearly related with  the  level of practice of the runners.  Thus M8  is the only trail runner   and improved his personal record on that occasion; he practices more  and developed a very uneven way  of running. 
 Table \ref{tab1}  shows that the values of  $c_1 (p) $ do not notably differ for different values of  $p$ and, when computed on a primitive of the signal,  are  shifted by 1.  We are in the situation described in Prop. \ref{propae}   and we conclude in the absence of oscillating singularities at almost every point.  
  This  result   also shows  that $\cu$, which may depend on the value of $p$ (see  \cite{Bandt2015} where it is shown  that it is the case for  lacunary wavelet series),  is intrinsic for such data.  We will see in Section \ref{MultivariateMar} that a bivariate analysis allows to investigate further in the nature of the pointwise singularities of the data.

 \begin{figure}
     \centering
     \includegraphics[scale=0.12]{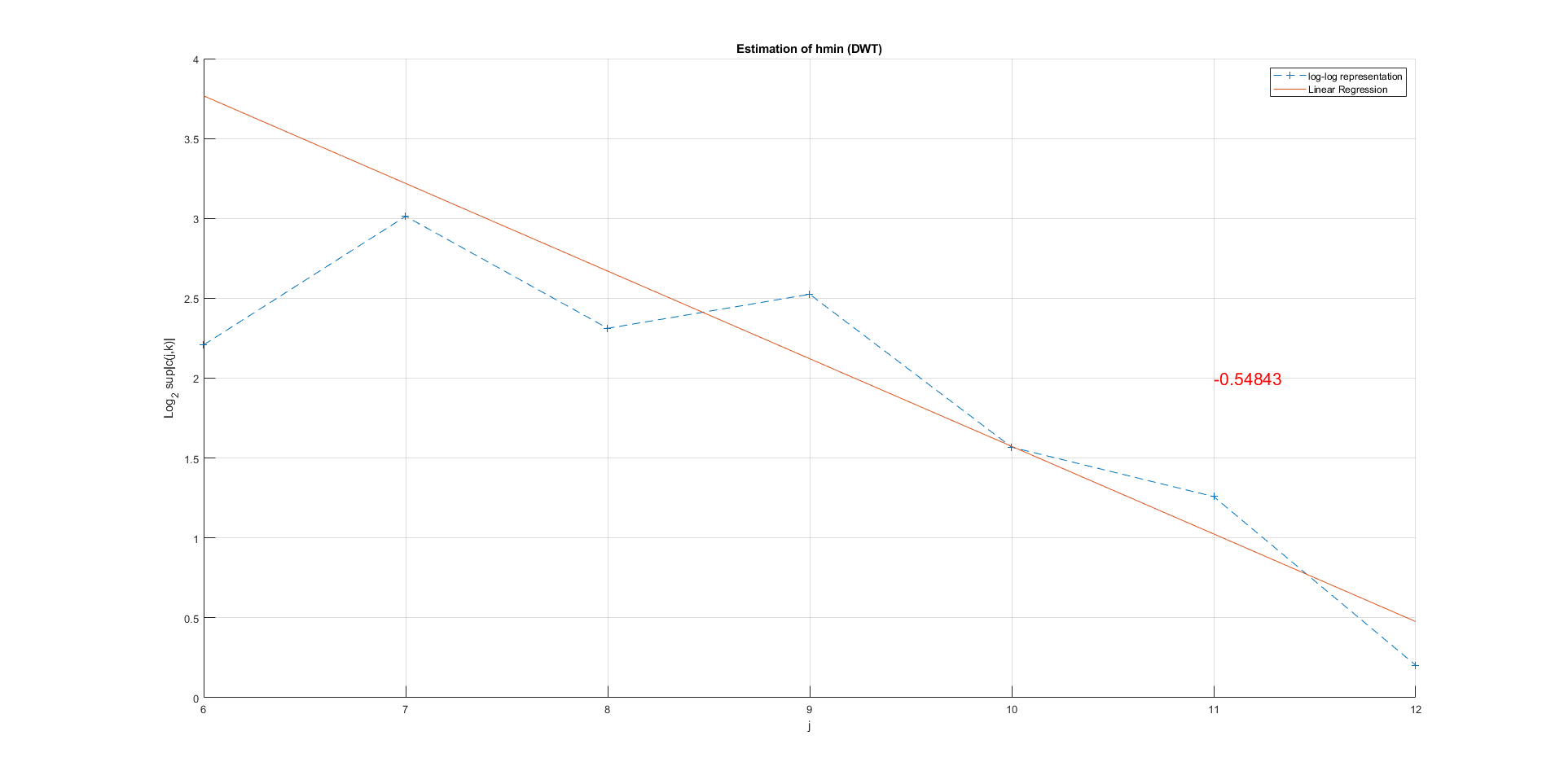}\qquad
     \includegraphics[scale=0.12]{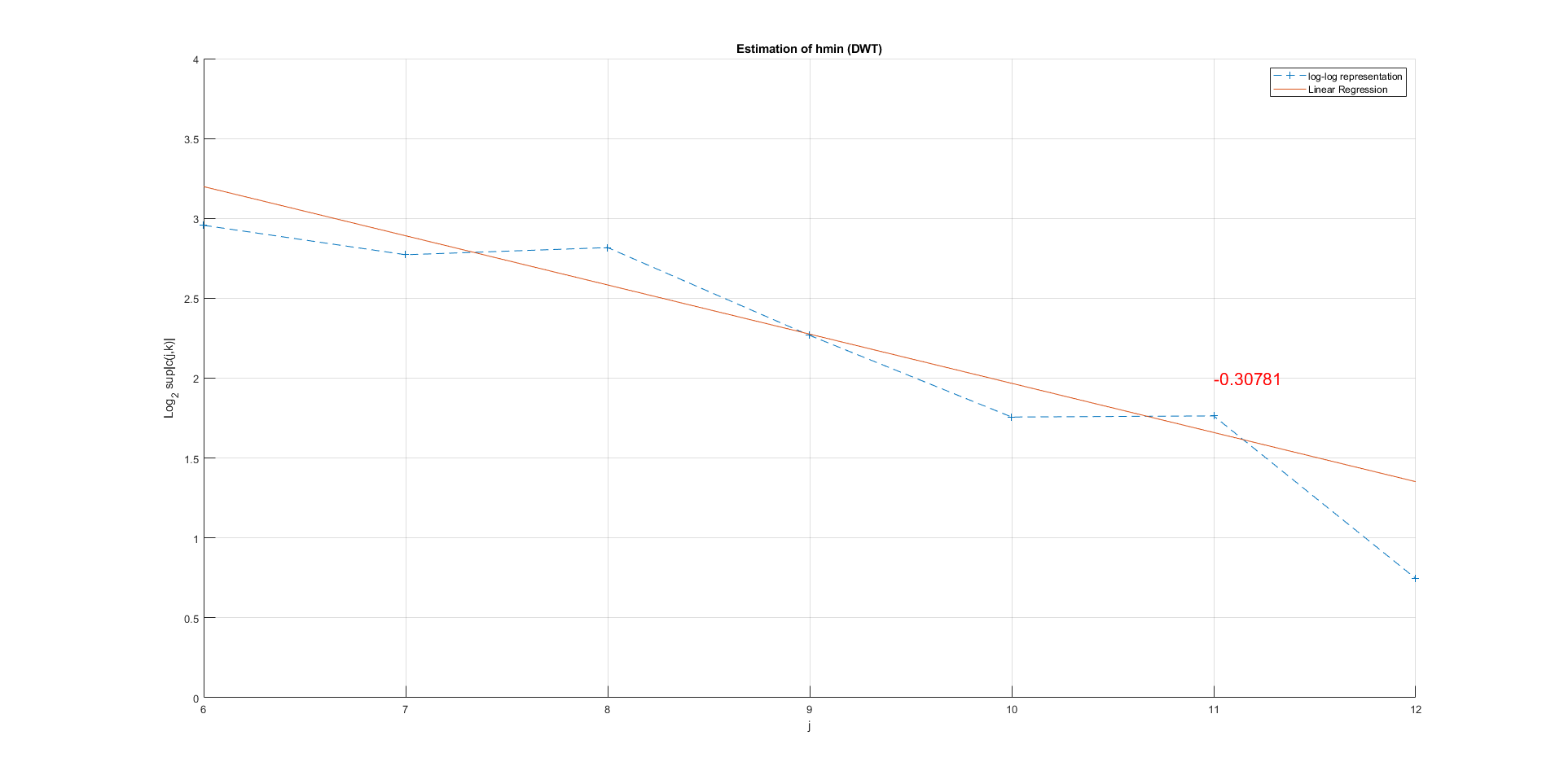}
     \caption{Estimation of $\Hmin$ by log-log regression for the heart rate of a marathon runner at the beginning (50\% first part of the race) on the left and the end (25\% last part of the race) on the right. The clear difference of the values obtained shows that the exponent $\Hmin$ is well fitted  to characterize the evolution of physiological rythms during the race. These data, together with the evolution of the parameter $c_1 (p)$,   are collected in Fig.\ref{fig:my_label} with $p=1$.}
     \label{hmindebfin}
 \end{figure}

\begin{figure}
    \centering
    \includegraphics[scale =0.35]{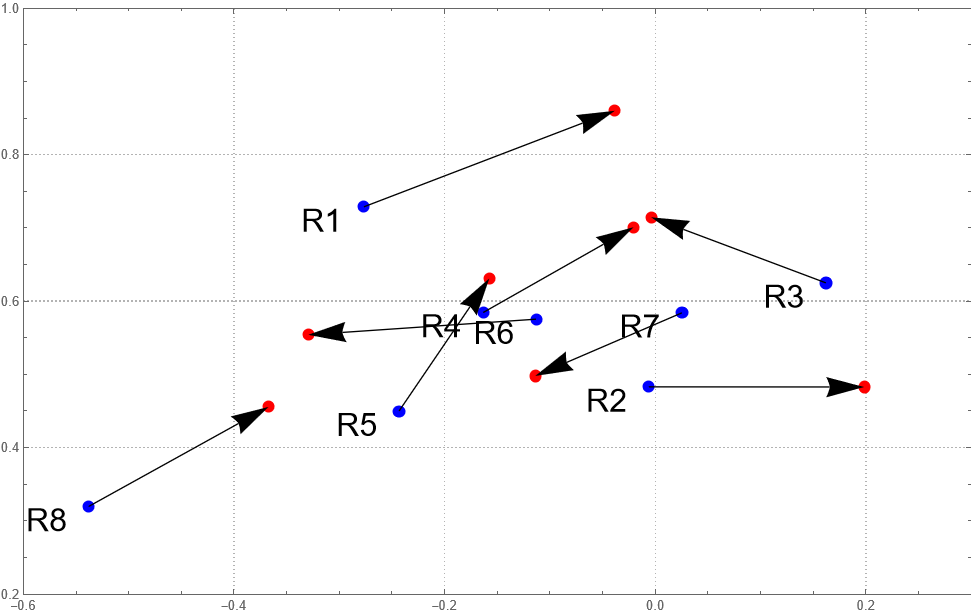}
    \caption{Evolution of the couple $(H_{min}, c_1(p))$ with $p=1$ deduced from the 1-spectrum of the heart rate between the beginning (in blue) and the end (in red) of the marathon: the evolutions are similar except for three runners: R3 and R6 who had great difficulties and R7 who is the least experienced runner with a much longer running time.}
    \label{fig:my_label}
\end{figure}

 We now consider the evolution of the multifractality parameters during a marathon:   at about  the  25th Km  (circa 60 \% of the race)
 runners feel an increased penibility  on the 
 RPE Borg  scale. Therefore we expect to find two regimes with different parameters  before and after this moment. This is put in evidence by
 Fig. \ref{fig:my_label}  which shows the evolution of the multifractality parameters during the first half and the last fourth of the  marathon thus putting in evidence  the different 
  physiological reactions  at about the  28th Km.  
    From the evolution of the multifractal parameters between the beginning and the end of the marathon race, we can distinguish between  the less experimented marathon runners, whichever their level of fitness, and those who know how to self pace their race. Indeed, according to the evolution of the couple $(\Hmin, c_1 (p))$, the less experimented (R 7) loosed the regularity of his heart rate variation. This shows that the mararathon running experience allows to feel how to modulate the speed for a conservative heart rate variability. From the evolution of the multifractal parameters between the beginning and the end of the marathon race, we can distinguish  between the less experimented marathon runners, whichever their level of fitness and  those who know how to self pace their race.  In \cite{pycke2022marathon} its was shown  that the  best marathon performance was achieved with a speed variation between extreme values. Furthermore, a phsyiological steady state (heart rate and other cardiorespiratory variables), are obtained with pace variation \cite{BillatPet}. This conclusion is in opposition with  the less experimented runners beliefs  that the constant pace is the best, following the mainstream non scientific basis recommendations currently available on internet.

   In Section \ref{MultivariateMar} we will investigate  the additional information  which is revealed by the joint analysis of several physiological  data.

\section{Multivariate multifractal analysis} 

\label{Multivariate}

 Up to now,  in  most applications, multifractal analysis was performed in univariate settings, (see a contrario \cite{Lux}), which was  mostly due to a lack of theoretical foundations and practical analysis tools. Our purpose in this section is to provide a comprehensive survey of the recent works that started to provide these foundations, and to emphasize the mathematical questions which they open. In particular,  multivariate spectra also encode  on specific data construction mechanisms.  
Multivariate multifractal analysis deals with the joint multifractal analysis of several functions.  For notational  simplicity, we assume in the following that we deal with two functions $f_1$ and $f_2$ defined on $\RR^d$ and that, to each function is associated a pointwise regularity exponent $h_1 (x) $ and $h_2 (x)$ (which need not be the same).

\subsection{Multivariate spectrum} 

\label{MultivariateSpec}

On the mathematical side, the main issue is to understand how the isoregularity sets
\[ E_{f_1} (H_1) = \{ x: h_1(x) = H_1 \} \quad \mbox{ and } \quad E_{f_2} (H_2) = \{ x: h_2(x) = H_2 \} \]
of each function are ``related''. A natural way to  translate this loose question into a precise mathematical problem is to ask for the  determination of the { \em multivariate multifractal spectrum} defined as the two-variables function
\BE { \cal D}_{(f_1,f_2)} (H_1, H_2) = \dim (\{ x: h_1(x) = H_1  \mbox{ and }  h_2(x) = H_2 \} ). \EE
this means that we   want to determine the dimension of the intersection of the two isoregularity sets $E_{f_1} (H_1) $ and $E_{f_2} (H_2) $.  The determination of  the dimension of the intersection  of two fractal sets  usually is a  difficult mathematical question, with  no general results available, and it follows that few multivariate spectra have been determined mathematically, see e.g.   \cite{Barr,Schmel} for  a joint analysis of invariant measures of dynamical systems.
One can  also mention correlated and anticorrelated binomial cascades, see Section \ref{BrwonianMT} for the definition of these cascades, and \cite{PRSA} for the determination of  bivariate spectra when two of these cascades are considered jointly.

On the mathematical side,   two types of results often show up. 
A first category follows from   the intuition supplied by intersections of smooth manifolds: 
In general, two surfaces in $\RR^3$  intersect along a curve and, more generally,   in $\RR^d$,  manifolds intersect \emph{generically} according to the \emph{sum of codimensions rule}: 
\[  \dim  (A \cap B) =  \min (\dim A  + \dim B -d, -\infty) \]
  (i.e. the ``codimensions'' $d-\dim A $ and $d- \dim B$ add up except if the output is negative, in which case we obtain the emptyset). 
This formula is actually valid for numerous examples of fractal sets, in particular when the Hausdorff and Packing dimensions of one of the sets $A$ or $B$ coincide (e.g. for general Cantor sets) \cite{Matti}; in that case ``generically'' has to be understood in the following sense: For a subset of positive measure among  all rigid motions $\sigma$, 
$  \dim  (A \cap \sigma (B) ) =  \min (\dim A  + \dim B -d, -\infty)$.
However the coincidence of Hausdorff and Packing dimensions needs not be satisfied by  isoregularity sets, so that such results  cannot be directly applied  for many mathematical models. 
The only result that holds in all generality is the following:
if $A$ and $B$ are two Borel subsets of $\RR^d$, then, for a generic set of  rigid motions $\sigma$, $ \dim(A \cap \sigma (B) ) \geq \dim A + \dim B -d$.
This leads to a first rule of thumb for multivariate multifractal spectra: 
When two functions are randomly shifted,  then  their singularity sets will be in ``generic'' position with respect to each other,  yielding 
\[ { \cal D}_{(f_1,f_2)} (H_1, H_2)    \geq  { \cal D}_{f_1} (H_1) +   { \cal D}_{f_2} (H_2) -d.  \] 
In practice, this result suffers from two limitations: the first one is that, usually,  one is not interested in randomly shifted signals but on the opposite for particular configurations where we expect the conjunction of singularity sets to carry relevant information.  Additionally,  for large classes of fractal sets, the {\em sets with large intersection}, the codimension formula is not optimal   
as they satisfy  
\[  \dim  (A \cap B) = \min ( \dim A , \;  \dim B ).  \] 
While  this alternative formula may seem counterintuitive,  general frameworks where it holds were uncovered, cf. e.g., \cite{Fal93,Dur,BaSeurBraz}  and references therein. 
This is notably commonly met  by { \em limsup sets}, obtained as follows:
There exists a collection of sets $A_n$ such that 
$A$
 is the set of points that belong to an infinite number of the $A_n$.
 This is particularly relevant for  multifractal analysis where the  singularity sets 
   $E_{H}^-$ defined in  (\ref {uppersets}) often turn out to be   of this type: 
 It is the case for   L\'evy processes or   random wavelet series, see e.g. \cite{Jaf6,AJ02,{JaffLev}}).
For multivariate multifractal spectra, this leads to an alternative formula 
\BE  \label{largint2}  { \cal D}_{(f_1,f_2)}  =   \min ( { \cal D}_{f_1} (H_1),  { \cal D}_{f_2} (H_2))  \EE 
expected to hold in competition with the codimension formula, at least for the sets $E_{H}^-$.
The existence of two well motivated  formulas in competition  makes it hard to expect that  general mathematical results could hold under fairly reasonable assumptions.    Therefore, we now turn towards the construction of multifractal formalisms adapted to a multivariate setting, first in order to inspect if this approach can yield  more intuition on the determination of multivariate spectra and, second, in order to derive new multifractality parameters which could be used for model selection and identification, and also in order to get some understanding on the ways that singularity sets of  several functions are correlated.

In order to get some intuition in that direction, it is useful to  start with a  probabilistic interpretation of the multifractal quantities that were introduced in the univariate setting.

\subsection{Probabilistic interpretation of scaling functions} 

\label{MultivariateWSF}

We consider the following probabilistic toy-model:  We assume that, for a given $j$,  the wavelet coefficients $(c_{ j,k})_{k\in \ZZ}$  of the signal considered share a common law $X_j$ and  display  short range memory, i.e.  become quickly decorrelated when   the wavelets $\psi_{j,k}$ and  $\psi_{j,k'}$ are located far away (i.e.  when $k-k'$ gets large);   then, the  wavelet structure functions
(\ref{equ-WSF}) can be interpreted as an empirical estimation of $\EEE ( |X_j |^p)$, i.e. the moments of the random variables  $X_j  $, and the  wavelet scaling function  characterizes the power law behaviour of these moments (as a function of the scale $2^{-j}$). This interpretation is  classically acknowledged for signals which display some stationarity, and the vanishing moments of the wavelets  reinforce this decorrelation even if the initial process displays long range correlations, see e.g. the studies performed on classical models such as fBm (\cite{agf95} and ref. therein).  We will not discuss the relevance of this model; we just note that his interpretation has the advantage of pointing towards   probabilistic tools when one shifts from one to several signals, and these tools will allow to introduce natural classification parameters which can then be used even when the probabilistic assumptions which led to their introduction  have no reason to hold. 

From now on, we  consider   two signals $f_1$ and $f_2$ defined on $\RR$ (each one satisfying the above assumptions) with wavelet  coefficients  respectively $c^1_{ j,k}$ and $c^2_{ j,k}$. The ``covariance'' of the wavelet coefficients at scale $j$  is estimated by the empirical {  correlations }  
\BE \label{crosscorr} \mbox{ for }  m,n  = 1,2,  \qquad S_{m,n} (j) = 2^{-j} \displaystyle\sum_{  k }   c^m_{ j,k} c^n_{ j,k}. \EE 
Log-log regressions of these quantities (as a function of  $\log (2^{-j})$ allow to determine if some power-law behaviour of these auto-correlations (if $m =n$)  and cross-correlations (if $m \neq n$)  can be put in evidence: When these correlations are found to be significantly non-negative, one  defines the { \em scaling exponents}  $H_{m,n}$ implicitly by 
\[    S_{m,n} (j)\sim 2^{-H_{m,n} j} \]
in the limit of small scales. Note that, if $m=n$, the exponent associated with the auto-correlation simply is  $\eta_f (2)$ and is referred to as the { \em Hurst exponent}  of the data.

Additionally, the { \em wavelet coherence function}  is defined as 
\[ C_{1,2} (j) = \frac{S_{1,2} (j)}{\sqrt{S_{1,1} (j) S_{2,2} (j) }}. \]
It ranges within the interval $[-1, 1]$ and quantifies, as a scale-dependent correlation coefficient, which scales
are  involved in the correlation of the two signals, see \cite{CRAS2019, Whitc}.

  Note that probabilistic denominations  such as ``auto-correlation'', ``cross-correlations''  and ``coherence function''  are used even if no probabilistic model is assumed, and used in order to derive scaling parameters obtained by log-log plot regression which can prove powerful  as  classification tools. 

\begin{figure}
    \centering
    \includegraphics[scale=0.3]{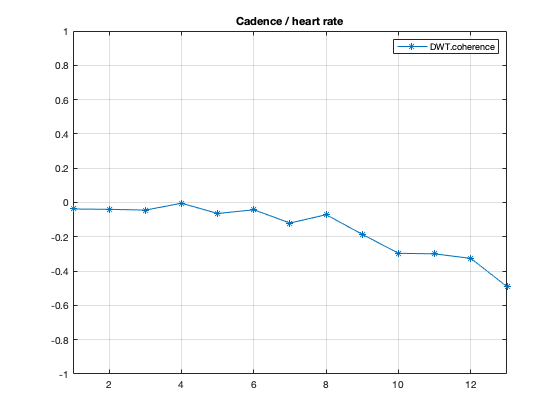} \quad
    \includegraphics[scale=0.3]{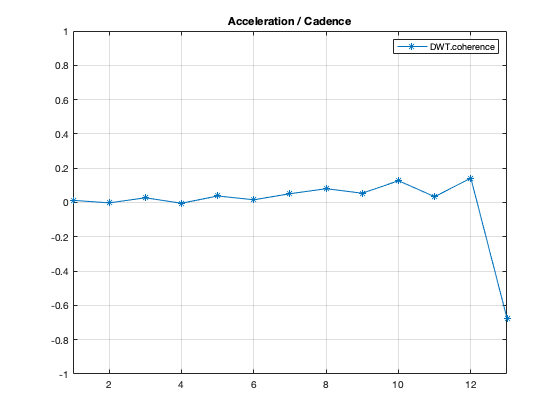} 
    \caption{  Wavelet coherence between heart-beat frequency and cadence (left) and  between acceleration and cadence (right).}
    \label{CrossCor}
\end{figure}

As an illustration, we estimated these crosscorrelations   concerning the following couples of data recorded on  marathon runners:    heart-beat frequency  vs.   cadence,  and  cadence vs. acceleration, see Fig. \ref{CrossCor}. In both cases, no correlation between the wavelet coefficients at a given scale  is put in evidence. Therefore, this is   a situation where  the additional bonus brought by measuring multifractal correlations is needed. Indeed, if  the cross-correlations of the signals do not  carry substantial information, this does not imply that the singularity sets of each signal are not related (as  shown  by the  example  supplied by { \em Brownian motions in multifractal time}, see  below in Section \ref{BrwonianMT}).  In that case, a natural idea is to look for correlations that would be revealed by   the multiscale quantities associated with pointwise exponents rather than by wavelet coefficients.

\subsection{Multivariate multifractal formalism} 

\label{MultivariateMF}

The idea that leads to a multivariate multifractal formalism is quite similar as the one which led us from wavelet scaling functions to leaders and $p$-leaders scaling functions: One should incorporate in the cross-correlations the multiscale quantities which allow to characterize pointwise regularity, i.e. replace wavelet coefficients by wavelet leaders in \eqref{crosscorr}. 

     Suppose that two pointwise regularity exponents $h_1  $ and $   h_2  $ defined on $\RR$ are given. We assume that each of these exponents  can be derived from  corresponding multiresolution quantities $d^1_{j,k}$,  and $d^2_{j,k}$ according to \eqref{carachqf}.
A {\em grandcanonical multifractal formalism} allows to estimate the {joint spectrum} ${ \mathcal D}(H_1,  H_2)$ of the couple  of exponents $(h_1  , h_2)  $ as proposed in  \cite{Meneveau90}. In the general setting provided by multiresolution quantities, it is derived as follows:
The {\em multivariate structure functions}   associated with  the couple  $(d^1_{j,k},   d^2_{j,k})$  are defined by
 \begin{equation}
 \label{equ-WSF2} \forall  r= (r_1, \ r_2) \in \RR^2, \qquad
 S (r,j) = 2^{-j} \displaystyle\sum_{ k }  ( d^1_{j,k} )^{r_1}  ( d^2_{j,k} )^{r_2},
 \end{equation}
 see \cite{Porqu2017,Benslim} for the seminal idea of proposing such  multivariate multiresolution quantities  as building blocks  of a { \em grandcanonical formalism}. 
 Note that they are  defined as a cross-correlation, which would be based on the quantities $ d^1_{j,k} $ and $ d^2_{j,k} $,  with the extra flexibility of raising them to arbitrary powers, as is the case for univariate structure functions.    The  corresponding {\em bivariate scaling function} is
  \BE \label{defscalond3}
  \zeta (r) =   \displaystyle \liminf_{j \rightarrow + \infty} \;\; \frac{\log \left( S (r,j)  \right) }{\log (2^{-j})}.  \EE
The {\em bivariate  Legendre spectrum}  is obtained through a 2-variable Legendre transform
\begin{equation}
\label{formult1}  \forall H = (H_1,  H_2) \in \RR^2, \qquad
{ \mathcal L}  (H) = \inf_{r\in \RR^2  } (1-\zeta (r) + H \cdot r),
\end{equation}
where $H\cdot r$ denotes the usual scalar product in $\RR^2$.
Apart from  \cite{Meneveau90},  this formalism has been  investigated in a  wavelet framework for joint H\"older and oscillation exponents in \cite{ABJM},   in an abstract general framework in \cite{pey2004},   and on wavelet leader and $p$-leader based quantities in \cite{Bandt2015,Porqu2017}.
\vspace{0.5cm}

{ \bf Remark:} The setting supplied by orthonormal wavelet bases is well fitted to  be extended to the multivariate setting, because the multiresolution quantities $d_\la$ are defined on a preexisting (dyadic) grid, which is shared by both quantities. Note that this is not the case for the WTMM, where the multiresolution quantities are defined  at the local maxima of the continuous wavelet transform (see \eqref{foncechbis}), and these local maxima differ for different signals; thus, defining multivariate structure functions in this  setting  would lead to the complicated questions of matching these local maxima correctly in order to construct bivariate structure functions similar to \eqref{equ-WSF2}.
\vspace{0.5cm}

The multivariate multifractal formalism is backed by only  few mathematical results. A first reason is that, as already mentioned,  the Legendre spectrum does not yield in general an upper bound for the multifractal spectrum, and this property is of key importance in the univariate setting. Another  drawback is that, in constradistinction with the univariate case, the scaling function \eqref{defscalond3}  has no function space interpretation. It follows that  there exists no proper setting for genericity results except if one defines a priori this function space setting (as in \cite{Benslim,Moez} where generic results are obtained  in  couples of function spaces endowed with the natural   norm on a product space). We meet here once again the problem of finding a ``proper'' genericity setting that would be fitted to the quantities supplied by scaling functions. 
We now list several positive results concerning multivariate Legendre spectra.

The following result of \cite{ACHA2018}  shows how to recover the univariate Legendre spectra from the bivariate one.

\BP \label{propbivleg} Let $d^1_{j,k}$,  and $d^2_{j,k}$  be two multiresolution quantities   associated with two pointwise exponents $h_1 (x) $ and $  h_2 (x) $.  The  associated uni- and  bi-variate  Legendre spectra are related as follows:  
\[ {\mathcal L}_1 (H_1)  = \sup_{H_2} \; {\mathcal L} ( H_1, H_2)  \quad \mbox{and} \quad  {\mathcal L}_2 (H_2)  = \sup_{H_1}\; {\mathcal L} ( H_1, H_2).    \]
\EP

This property implies  that results similar to  Theorem  \ref{theo1}  hold in the multivariate setting.

 \BC  \label{coro1} Let $d^1_{j,k}$  and $d^2_{j,k}$  be two multiresolution quantities  associated with two pointwise exponents $h_1 (x) $ and $  h_2 (x) $.    The following results on the couple $(h_1(x), h_2(x)) $ hold:
 \begin{itemize}
 \item If the bivariate Legendre spectrum has a unique maximum  for  $(H_1, H_2)= (c_1,c_2)$, then 
\BE \label{almoster2} \mbox{ for almost every } x, \qquad h_1 (x) = c_1 \quad \mbox{ and  } \quad h_2 (x) = c_2 . \EE 
\item If the leader scaling function is affine then  
  \[ \exists (c_1, c_2),  \quad \forall x, \qquad h_1 (x) = c_1 \quad \mbox{ and  } \quad h_2 (x) = c_2.\] 
\end{itemize}
\EC 

Note that  the fact that the leader scaling function is affine is equivalent to the fact that the bivariate Legendre spectrum is supported by a point. In that case, if the exponents $h_1$ and $h_2$ are associated with the functions 
$f_1$ and $f_2$, then they  are  monoh\"older functions.
\vspace{0.5cm}   

{ \bf Proof:} The first point holds because, if  the bivariate Legendre spectrum has a unique maximum, then,  its projections on the $H_1$ and the $H_2$ axes also have a unique maximum at respectively $H_1 = c_1$ and $ H_2= c_2$ and    Proposition \ref{propbivleg} together with Theorem  \ref{theo1} imply (\ref{almoster2}). 

As regards the second statement, one can use  Proposition \ref{propbivleg}: If the bivariate scaling function is affine, then ${\mathcal L} ( H_1, H_2)$ is supported by a point, so that  Proposition \ref{propbivleg} implies that it is also the case for univariate spectra ${\mathcal L} ( H_1)$ and ${\mathcal L} (H_2)$, and  Theorem  \ref{theo1} then implies that  $h_1$ is constant and the same holds for $h_2$.
\vspace{0.5cm}

Recall that, in general, the bivariate Legendre spectrum does not yield an upper bound for the multifractal spectrum (in contradistinction with the univariate case), see \cite{PRSA} where a counterexample is constructed; this limitation raises many open questions: Is there  another way to construct a Legendre spectrum which would yield an upper bound for ${ \cal D} (H_1, H_2)$? which information can actually be derived from the Legendre spectrum?  A first positive result was put in light in \cite{PRSA}, where  a notion of ``compatibility'' between  exponents is put in light and   is shown to hold for  several models: When this property holds, then  the  upper bound property is satisfied. It is not clear that there exists  a general way to check directly on the data if it is satisfied; however,   
 an important case where it is the case is when the exponents derived are the H\"older exponent and one of the ``second generation exponents'' that we mentioned, see \cite{Bandt2015,Porqu2017}. In that case, the upper bound property holds, and  it   allows to  conclude that the signal does not display  e.g. oscillating singularities, an important issue  both theoretical and practical. Let us mention a situation where this question shows up: In \cite{Balanca}, P. Balanca showed the existence of oscillating singularities in the sample of some L\'evy processes and also showed that they are absent in others (depending on the L\'evy measure which is picked in the construction); however, he only worked out several examples, and  settling the general case is an important issue; numerical  estimations of such bivariate spectra could help to make the right conjectures in this case.   

 The  general results listed in Corollary \ref{coro1}  did not require  assumptions on  correlations between the exponents $h_1$ and $h_2$.  We now investigate  the implications of such correlations on the joint Legendre spectrum. 
For that purpose,  let us come back to the probabilistic interpretation of the structure functions (\ref{equ-WSF2}) in terms of  cross-correlation of  the  $ ( d^1_{j,k} )^{r_1}$ and $  ( d^2_{j,k} )^{r_2} $.  
As in the univariate case, if we assume that, for a given $j$,  the multiresolution quantities  $d^1_{j,k} $ and $d^2_{j,k} $  respectively share  common laws $X^1_j$  and $X^2_j$ and  display  short range memory, then 
(\ref{equ-WSF2}) can be interpreted as an empirical estimation of $\EEE ( |X^1_j |^{r_1} |X^2_j |^{r_2})$.  If we  additionally assume that the $(d^1_{j,k}) $ and $(d^2_{j,k} )$ are independent, then we obtain 
\[  S (r,j) =  \EEE ( |X^1_j |^{r_1} |X^2_j |^{r_2})= \EEE ( |X^1_j |^{r_1} ) \cdot \EEE ( |X^2_j |^{r_2}) , \]
which can be written
   \BE \label{strcubiz}  S (r_1, r_2,j)    =  S^1 (r_1,j)  S^2 (r_2, j) . \EE
Assuming that $\liminf  $ in (\ref{defscalond3}) actually  is a limit, we obtain $S (r_1, r_2,j) \sim 2^{ -(\zeta^1(r_1)+\zeta^2(r_2)) j}$ yielding
  $  \zeta( r_1, r_2) =  \zeta^1 (r_1)  + \zeta^2 (r_2). $
Applying  (\ref{formult1}), we get
  \[ {\mathcal L}  (H_1, H_2 ) = \inf_{(r_1, r_2)\in \RR^2  } (1-\zeta^1( r_1) + \zeta^2( r_2) + H_1 r_1 + H_2 r_2) \]
 \[  = \inf_{r_1  } (1-\zeta^1( r_1) + H_1 r_1  ) + \inf_{r_2  }
 (1- \zeta^2( r_2) +  H_2 r_2) -1,  \]
 which  leads to 
 \BE \label{spectindep} {\mathcal L} (H_1, H_2) = {\mathcal L} (H_1) + {\mathcal L} (H_2) -1.  \EE
 Thus,  under stationarity and independence, the codimension rule applies for the multivariate Legendre spectrum. 
In practice, this means that any departure of the Legendre spectrum from (\ref{spectindep}), which can be checked on real-life data,  indicates that one of the assumptions required to yield (\ref{spectindep}) (either stationarity or independence) does not hold. 


\begin{figure}
    \centering
    \includegraphics[scale=0.3]{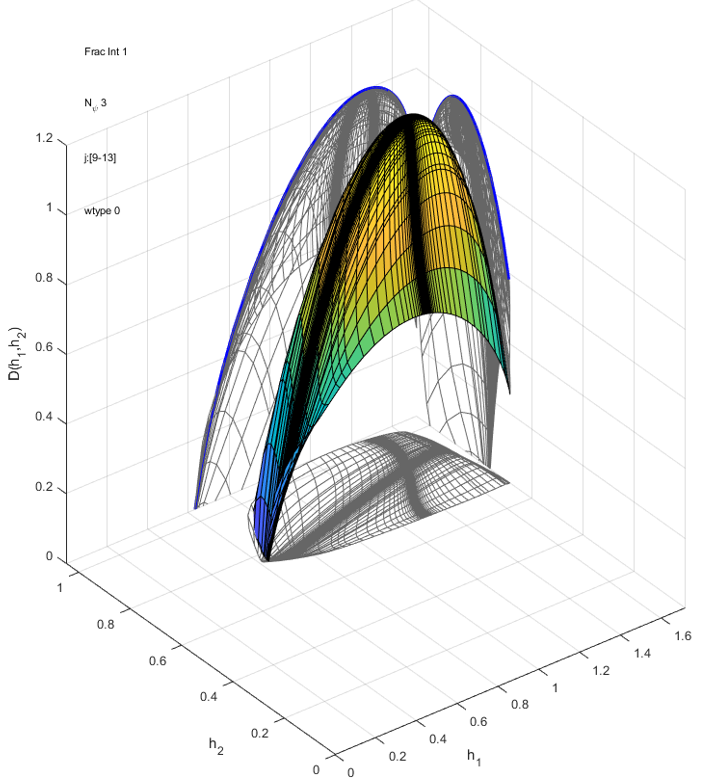} \quad
    \includegraphics[scale=0.3]{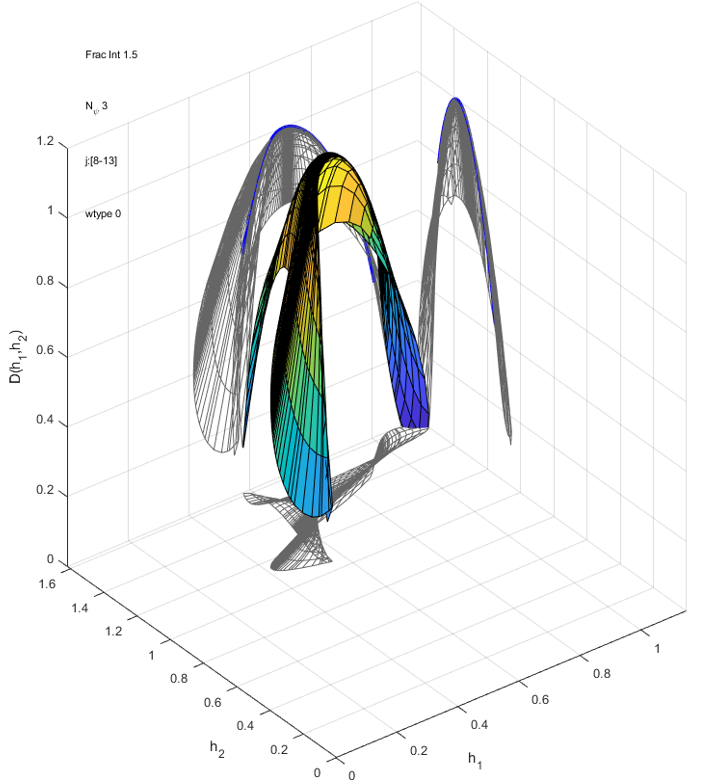}
    \caption{On the left, the bivariate multifractal spectrum between heart-beat frequency primitive and cadence primitive are shown,  and, on the right, the bivariate multifractal spectrum between acceleration and cadence with fractional integral of order 1.5 are shown. This demonstrates the  strong correlation between the pointwise singularities of the two data: indeed the bivariate spectra are almost carried by a segment, and a bivariate spectrum carried by a line $H_2 =  a H_1 + b$ indicates  a perfect match between the pointwise exponents accrording to  the same relationship: $\forall x$, $ h_2 (x) = a h_1(x) + b$ }
    \label{SpectBivar}
\end{figure}


\label{intfrac} 

\begin{figure}
    \centering
    \includegraphics[scale = 0.1]{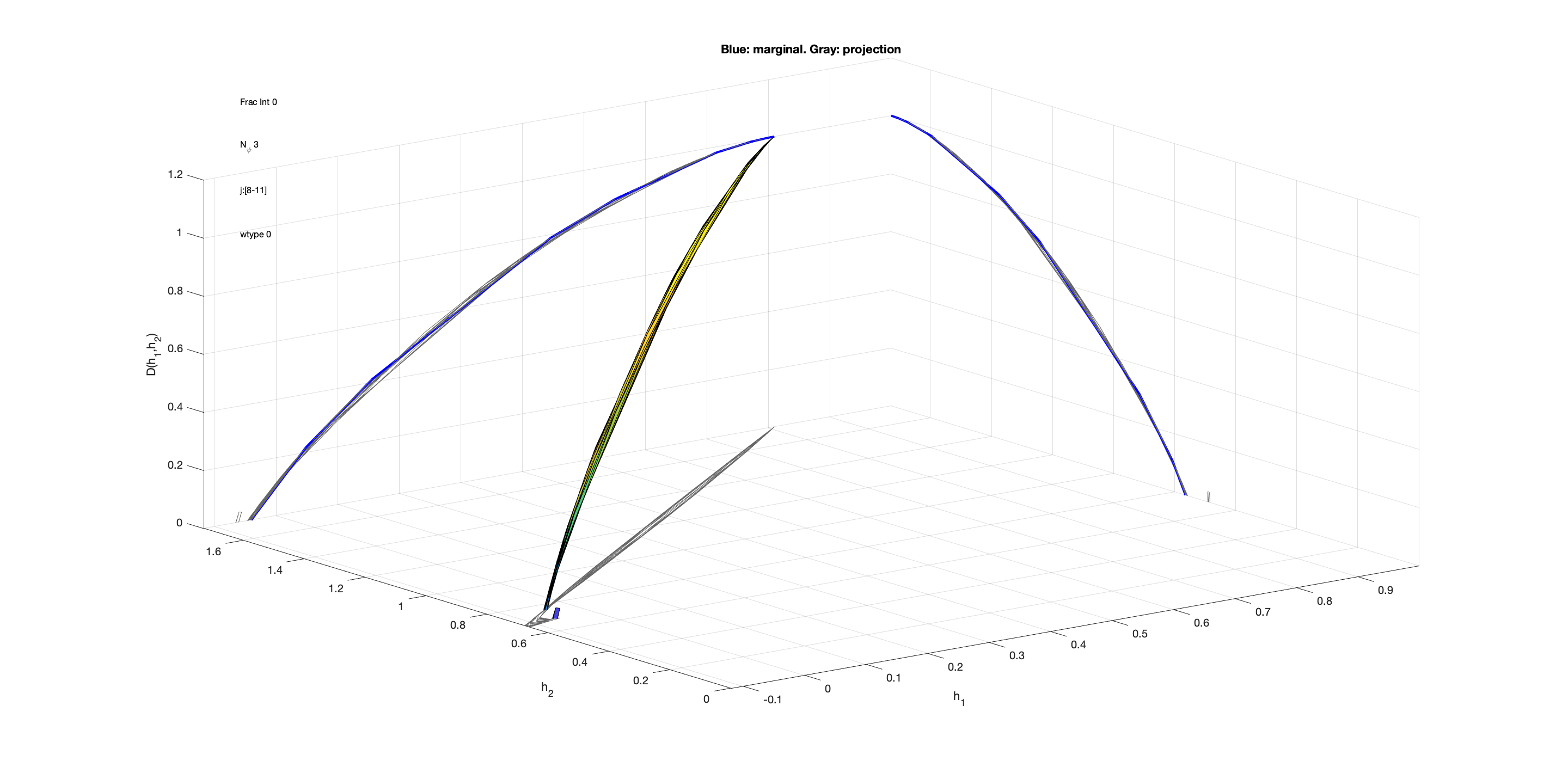}
    \caption{Bivariate $1$-spectrum of heartbeat frequency and its primitive: the bivariate spectrum lines up perfectly along the line  $H_2 = H_1 +1$. }
    \label{BivarFC}
\end{figure}

\begin{figure}
    \centering
    \includegraphics[scale = 0.1]{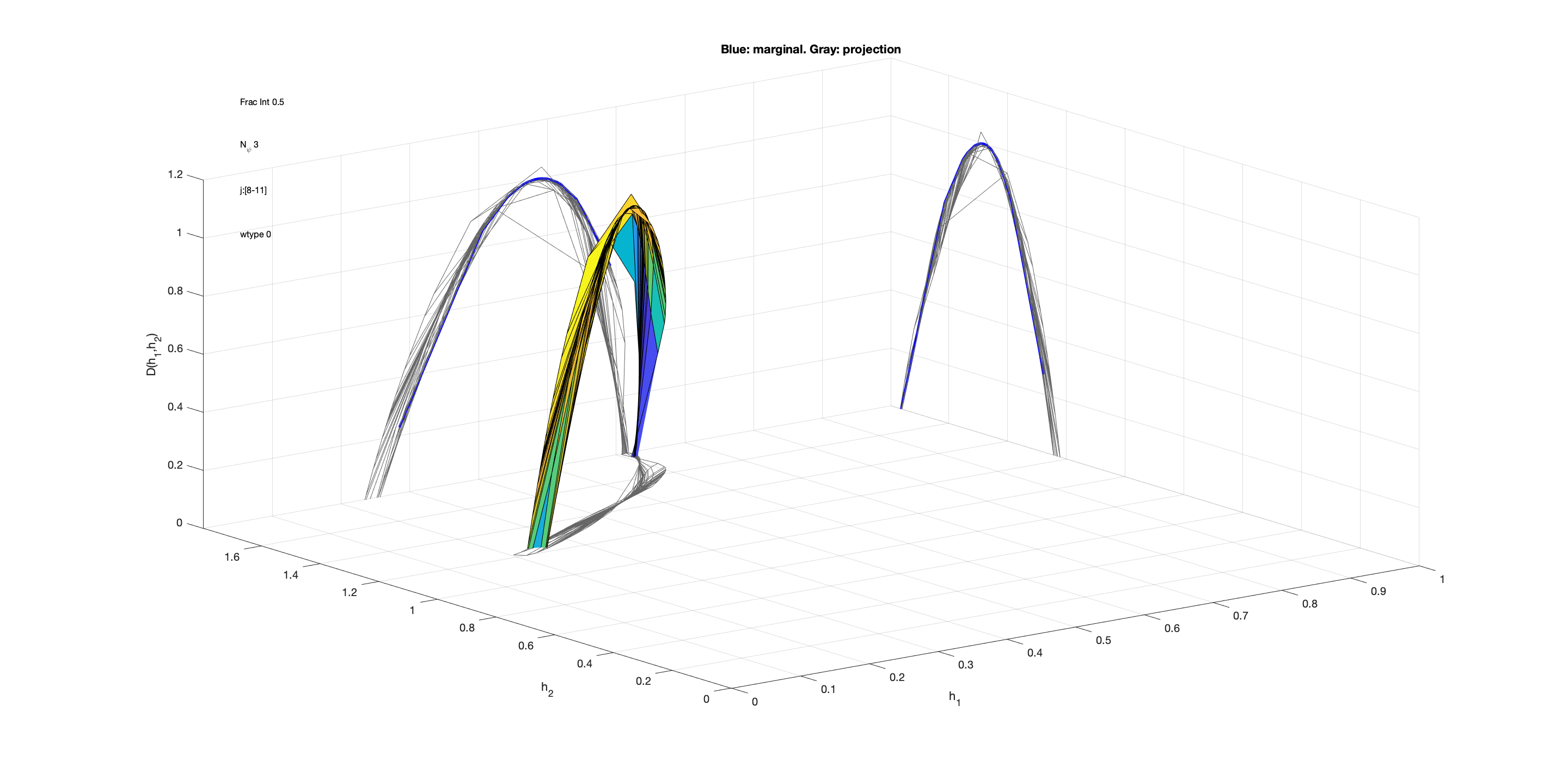}
    \caption{Bivariate  H\"older spectrum of fractional integrals order $1/2$ and $3/2$ of cadence: the bivariate spectrum lines up perfectly along the line  $H_2 = H_1 +1$.}
    \label{BivarCad}
\end{figure}

\begin{figure}
    \centering
    \includegraphics[scale = 0.1]{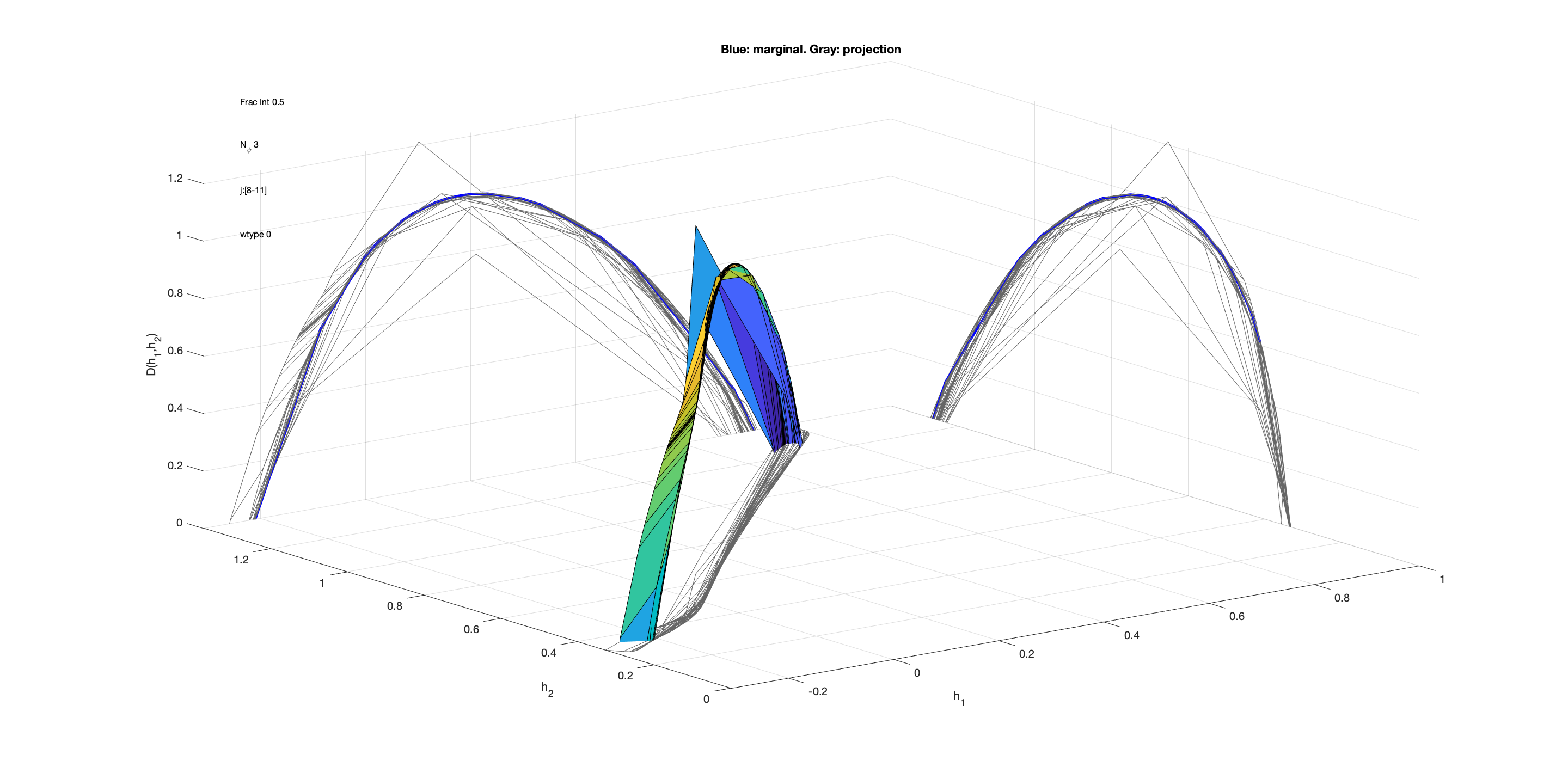}
    \caption{Bivariate  H\"older spectrum of fractional integrals of order $1/2$ and $3/2$ of acceleration: the bivariate spectrum lines up perfectly along the line  $H_2 = H_1 +1$.}
    \label{BivarAcc}
\end{figure}

As a byproduct, we now show that multivariate multifractal analysis can give information on the nature of the singularities of { \em one} signal, thus complementing results such as Proposition \ref{propae} which yielded almost everywhere information of this type. 
Let us consider the  joint multifractal spectrum of a function $f$ and its fractional integral of order $s$, denoted by  $f^{(-s)}$. If  $f$ only has  canonical  singularities, then the H\"older exponent of $f^{(-s)}$ satisfies $ \forall x_0$, $h_{f^{(-s)}}(x_0) = h_{f}(x_0) +s $, so that the  joint Legendre spectrum is supported by the line $H_2 = H_1 +s$. In that case, the synchronicity assumption is satisfied and one can conclude that the  joint multifractal spectrum  is supported by the same segment;  a contrario, a   joint Legendre spectrum which is not supported by this line is interpreted as   the signature  of { \em  oscillating singularities} in  the data, as shown by the discussion above concerning the cases where the upper bound for  bivariate spectra holds.    Figs. \ref{BivarFC}, \ref{BivarCad}  and \ref{BivarAcc} illustrate this use of bivariate multifractal analysis: In each case, a signal and its primitive are jointly analyzed: The three signals are collected on the same runner  and the whole race is analyzed.  Fig. \ref{BivarFC} shows the analysis of heartbeat,  Fig. \ref{BivarCad}  shows the cadence and   Fig. \ref{BivarAcc} shows the acceleration. In the first case, the analysis is performed directly on the data using a $p$-exponent with $p= 1$, whereas, for  the two last ones, the analysis is performed on a fractional integral of order $1/2$. In each case, the results yield a bivariate Legendre spectrum supported by the segment $H_2 = H_1 + s$, which confirms the almost everywhere results obtained in Section \ref{Marathon1}:  The data only contain canonical singularities.

\subsection{Fractional Brownian motions in multifractal time} 

\label{BrwonianMT}

In order to put in light the additional information between wavelet correlations and  bivariate scaling functions (and the associated Legendre spectrum), we consider the model supplied by Brownian motion in multifractal time, 
which  has been proposed by B. Mandelbrot  \cite{Mandelbrot1997,MandCal97} as a simple model for financial time series:  Instead of the classical Brownian model $B(t)$, he introduced  a time change (sometimes referred to as a { \em \hspace{-2mm}  subordinator}) 
\[ B(f(t))= (B \circ f)(t)  \]
  where the irregularities of $f$ model the fluctuations of the intrinsic ``economic time'', and  typically is a multifractal function.  In order to be a ``reasonable'' time change,  the function $f$ has to be continuous and  strictly increasing; such functions usually are obtained as distribution functions of probability measures $d\mu$ supported on $\RR$ (or on an interval), and  which have no atoms (i.e. $\forall a \in \RR$, $\mu (a) =0$);  typical examples are supplied by deterministic or random   cascades, and this is the kind of  models that  were advocated by B.  Mandelbrot in \cite{Mandelbrot1997}.   Such examples will allow to illustrate the different notions that we introduced, and the additional information which is put into light by the   bivariate Legendre spectrum and is absent  from wavelet correlations.

Let us  consider  the slightly more general setting of one  fBm of Hurst exponent $\al$ (the cas of Brownian motion corresponds to $\al = 1/2$) modified by a  time change $f$. In order to simplify its theoretical multifractal analysis, we take for pointwise regularity exponent  the H\"older exponent  and we  make the following assumptions of $f$: We assume that it has only canonical singularities and that, if they exist,  the non-constant terms of the Taylor polynomial of $f$ vanish at  every point even if the H\"older exponent at some  points is larger than 1 (this is typically the case for primitives of singular measures). 
In that case,  classical uniform estimates on increments of fBm, see \cite{KahCamb} imply that 
\BE \label{holponcbrown}  \mbox{ a.s. } \quad \forall t, \qquad h_{B\circ f} (t) = \alpha h_f (t),  \EE  
so that 
\[ \mbox{ a.s. } \quad \forall H, \qquad   { \cal D}_{B\circ f} (H) =  { \cal D}_{f} (H/ \al ) ; \]
Note that the simple conclusion \eqref{holponcbrown} may fail if the Taylor polynomial is not constant at every point, as shown by the simple example supplied by $f (x) = x$ on the interval $[0,1]$.

We now consider $B_1\circ f$ and $B_2\circ f$: two independent fBm   modified by the { \em same  deterministic time change}  $f$   (with the same assumptions as above). It follows from  (\ref{holponcbrown}) that, with probability 1,  the  H\"older exponents of $B_1\circ f$ and $B_2\circ f$ coincide everywhere, leading to  the following multifractal spectrum, which holds almost surely: 
\BE \label{specbivarbrown}    
\left\{ 
\begin{array}{rl}
 \mbox{ if } H_1 = H_2, &    { \cal D}_{(B_1\circ f,B_2\circ f )} (H_1, H_2 ) =  { \cal D}_{f} \left(\displaystyle\frac{H_1}{\al}\right)    \\ &    \\ 
  \mbox{ if } H_1 \neq  H_2, 
  &  { \cal D}_{(B_1\circ f,B_2\circ f )} (H_1, H_2 ) =  -\infty .      
\end{array}
\right.
\EE
  Fig.  \ref{fbmSpect}  gives a numerical backing of this result: The Legendre spectrum numerically obtained corresponds to the theoretical  multifractal spectrum.  Let us give a non-rigourous  argument which backs this result: The absence of oscillating singularities in the data  implies that the maxima in the wavelet leaders  are attained for a $\la'$ close to $\la$, so that  the wavelet leaders of a given magnitude will be close to coincide for both processes,  and  therefore 
the bivariate structure functions (\ref{equ-WSF2})  satisfy 

 \[  
 S_f (r,j) = 2^{-j} \displaystyle{\sum}_{ \la \in \La_j }  ( d^1_\la )^{r_1}  ( d^2_\la )^{r_2} \sim 
 2^{-dj} \displaystyle{\sum}_{ \la \in \La_j }  ( d^1_\la )^{r_1 + r_2}   
\] 
so that 
\[ \mbox{ a.s. }, \quad \forall r_1, r_2, \qquad \tilde{\zeta}   (r_1, r_2) =  \zeta (r_1 + r_2). \] 
where $\tilde{\zeta}$ is the bivariate scaling function of the couple $(B_1\circ f,B_2\circ f )$ and  ${\zeta}$ is the univariate scaling function of $B_1\circ f$. 
Taking a  Legendre   transform yields that the bivariate Legendre spectrum $ { \cal L} (H_1, H_2)$  also satisfies  a similar formula as 
(\ref{specbivarbrown}), i.e.   
\BE \label{specbivarbrown2}     \mbox{ a.s. }, \quad \forall H_1, H_2, 
\left\{ 
\begin{array}{rl}
 \mbox{ if } H_1 = H_2, &    { \cal L}_{(B_1\circ f,B_2\circ f )} (H_1, H_2 ) =  { \cal L}_{f} \left(\displaystyle\frac{H_1}{\al}\right)   \\ &  \\ 
  \mbox{ if } H_1 \neq  H_2, 
  &  { \cal L}_{f} \left(H_1, H_2\right)  =  -\infty .      
\end{array}
\right.
\EE

Let us now estimate  the  wavelet  cross correlations.  Since $f$ is deterministic, the processes $B_1\circ f $ and $B_2\circ f $ are two independent centered Gaussian processes. Their wavelet coefficients $c^1_{ j,k} $ and $ c^2_{ j,k}$ 
therefore are independent centered Gaussians, and,  at scale $j$  the quantity 
\[  \qquad S_{m,n} (j) = 2^{-j} \displaystyle\sum_{  k }   c^1_{ j,k} c^2_{ j,k}   \]
is an empirical estimation of their covariance, and therefore  vanishes (up to small statistical fluctuation). 
 In contradistinction with the bivariate spectrum, the { wavelet  cross correlations } reveal the decorrelation of the processes but does not yield information of the correlation of the singularity sets.

 In order to illustrate these results, we will use for time change the distributuon function of a    binomial cascade  $\mu_p$  carried on $[0,1]$. Let  $p \in (0,1)$;  $\mu_p$  is the only probability measure on $[0, 1]$   defined by recursion as follows: Let 
  $\la \subset [0, 1]$ be a dyadic interval of length $2^{-j}$; we denote by $\la^+$ and $\la^-$ respectively its two ``children'' of length $2^{-j-1}$, $\la^+$ being   on the left  and $\la^-$ being  on the right. Then, $\mu_{p}$  is the only probability measure carried by $[0, 1]$ and satisfying 
 \[ \mu_p (\la^{+}) = p \cdot\mu_p (\la) \quad \mbox{   and }  \quad \mu_p (\la^{-}) = (1-p) \cdot \mu_p (\la) . \]
 Then the  corresponding time change is the function 
 \[ \forall x \in [0, 1] \qquad  f_{\mu_p} (x) =  \mu_p ([0, x]). \]

 \begin{figure}
    \centering
    \includegraphics[scale = 0.15]{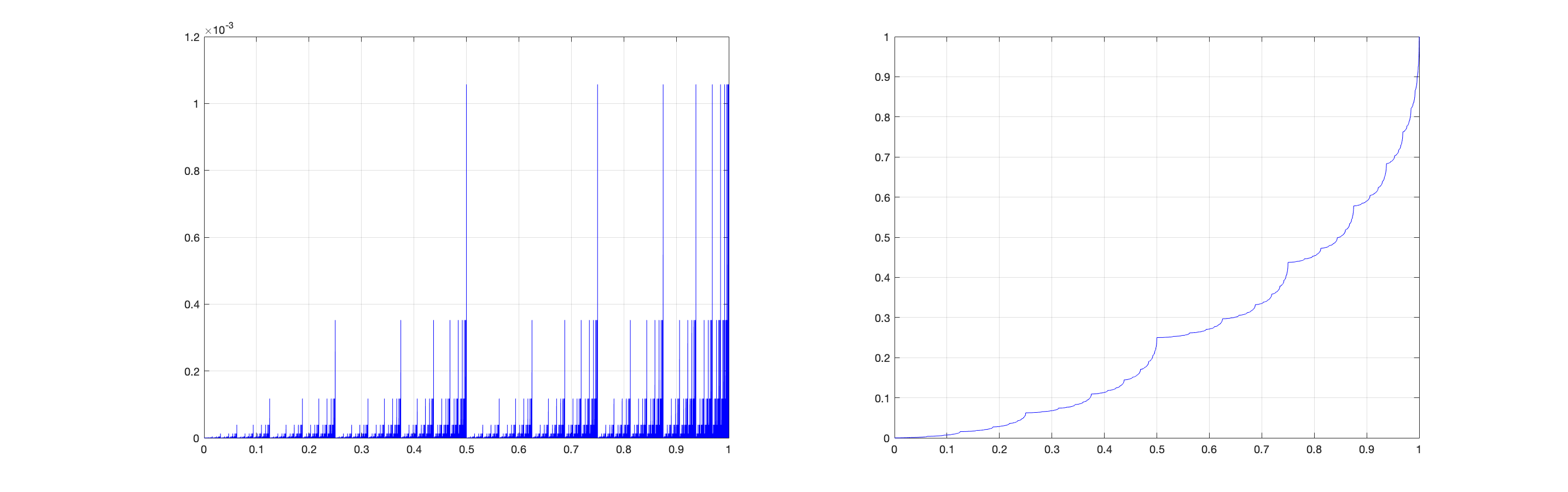}
    \caption{Binomial measure with $p=1/4$ (left) and  its repartition function (right) which is used as the time change in Fig.  \ref{fbmCT}.}
    \label{binom}
\end{figure}

 In Fig. \ref{binom},  we show the  binomial cascade $\mu_{1/4}$ and its distribution function, and in Fig.  \ref{fbmCT} we use this time change composed with a fBm of  Hurst exponent $\al=0.3$. 
 
\begin{figure}
    \centering
    \includegraphics[scale = 0.25]{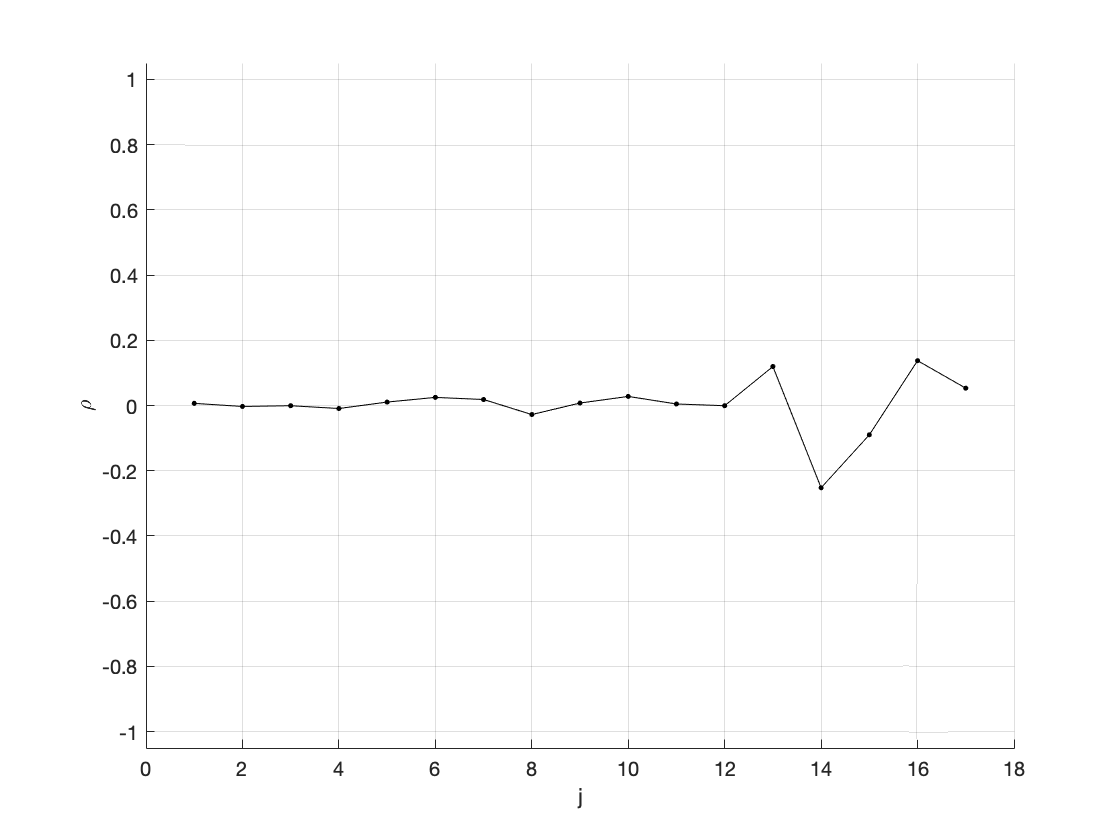}
    \caption{Cross-correlation of the wavelet coefficients of two independent  fBm  with the same time change : the distribution function of the  binomial measure $\mu_p$  with $p=1/4$. The  Cross-correlation  reflects the independence of the two processes.}
    \label{fbmCross}
\end{figure}

 \begin{figure}
    \centering
    \includegraphics[scale = 0.13]{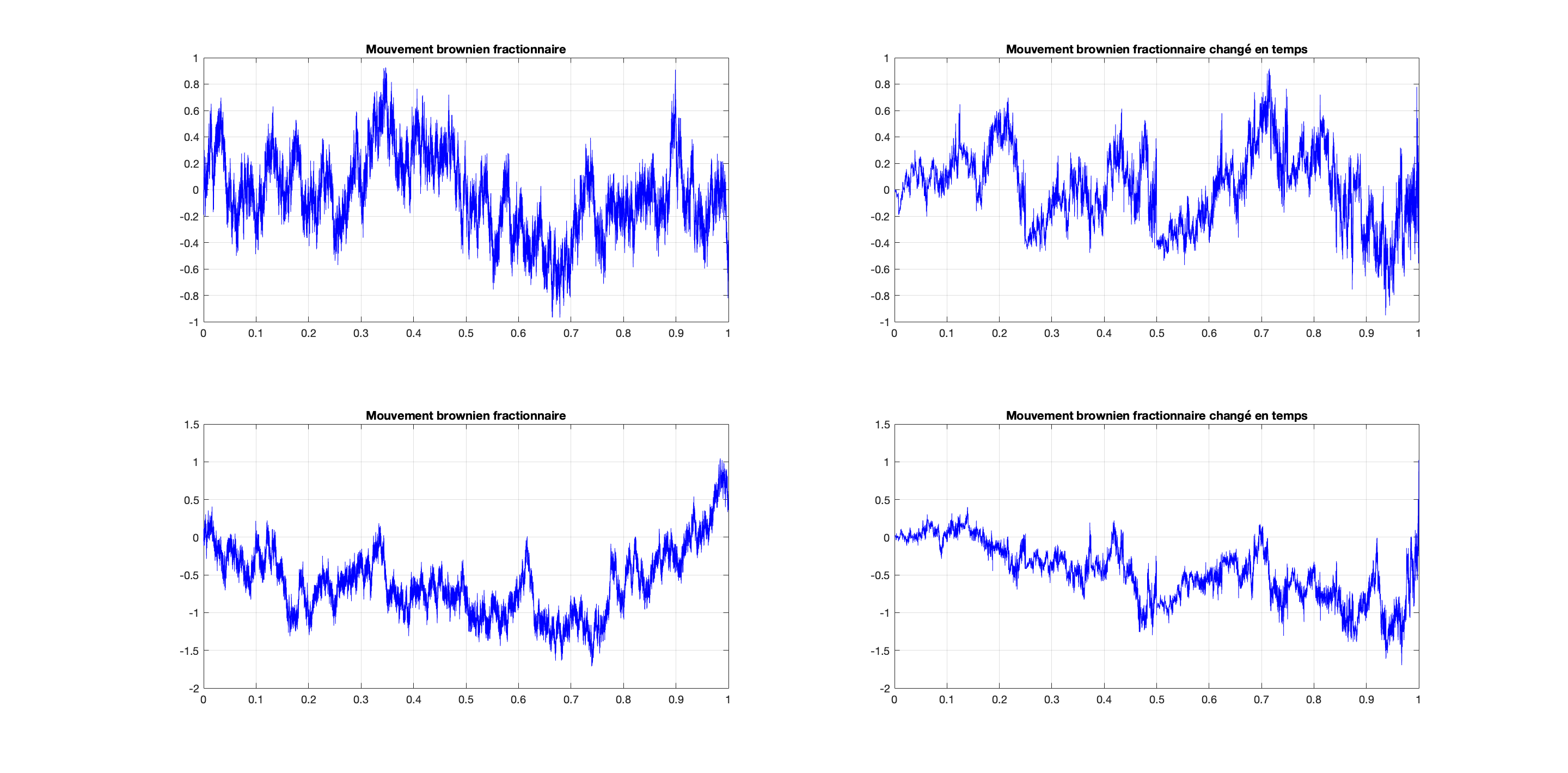}
    \caption{fBm with $H=0.3$ subordinated by the multifractal time change supplied by the distribution function of   the binomial measure $\mu_p$ with  $p=1/4$.}
    \label{fbmCT}
\end{figure}

{ \bf Remarks:} The fact  that  the same  time change is performed does not play a particular role for the estimation of the  wavelet  cross-correlations; the same result would follow for two processes $B_1\circ f $ and $B_2\circ g $ with $B_1$ and $B_2$  independent, and where $f$ and $g$ are two deterministic time changes. Similarly, $B_1$ and $B_2$ can be replaced by two (possibly different) centered Gaussian processes.

\begin{figure}
    \centering
    \includegraphics[scale = 0.13]{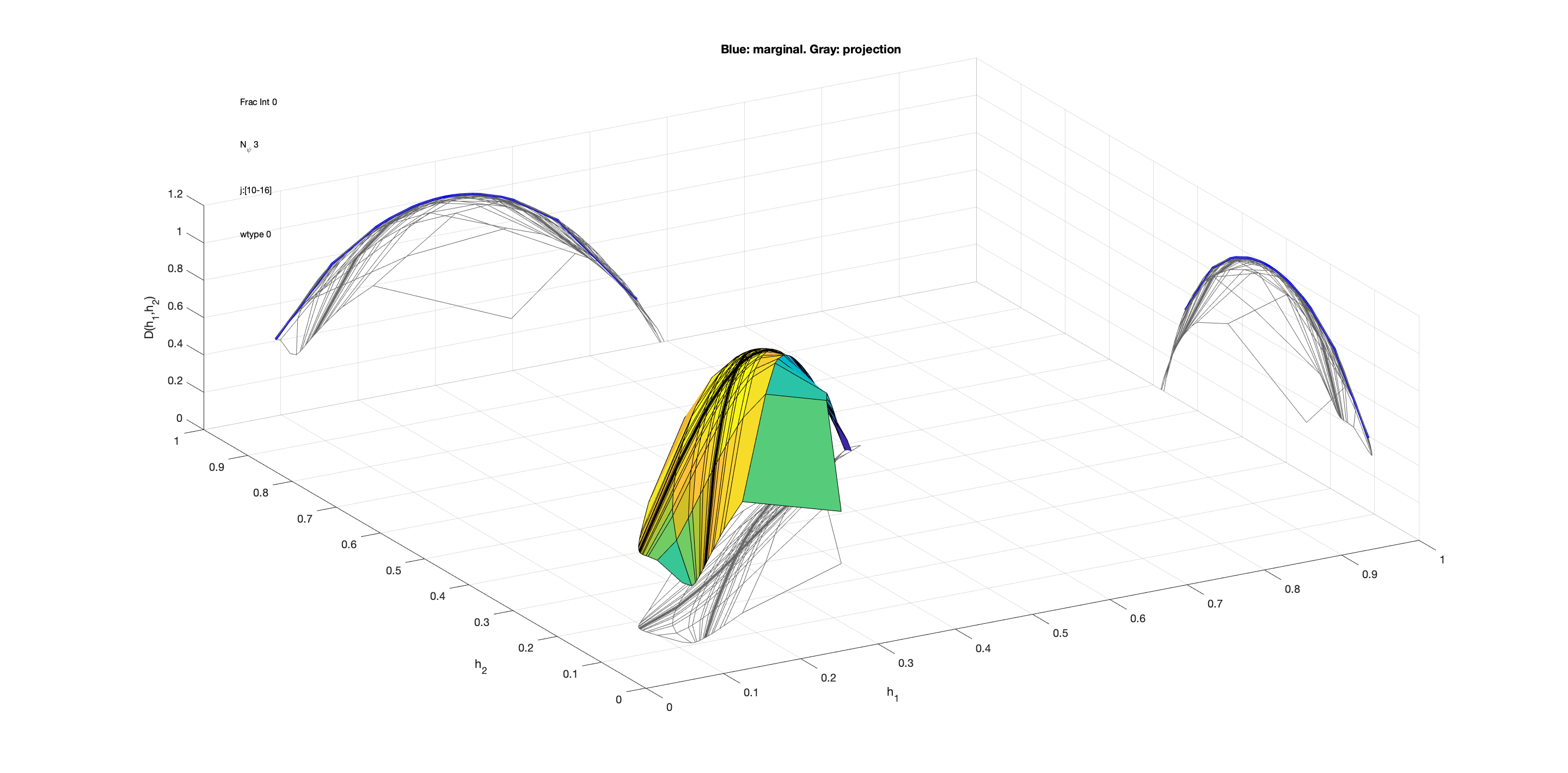}
    \caption{Bivariate multifractal spectrum of two  independent fBm with the same time  change: the distribution function of a   binomial measure  with $p=1/4$; in contradistinction with the cross-correlation of wavelet coefficients, the wavelet leaders are strongly correlated, leading to a bivariate  Legendre spectrum theoretically supported by the line $H_1 = H_2$, which is close to be the case numerically. }
    \label{fbmSpect}
\end{figure}
 
 Let us mention at this point that the mathematical problem of understanding what is the multifractal spectrum  of the composition $f\circ g$ of two multifractal functions $f$ and $g$, where $g$ is a { \em time subordinator} i.e. an increasing function,  is a largely open problem (and is posed  here in too much generality to find  a general  answer).  This problem was initially raised by B. Mandelbrot and  also investigated R. Riedi \cite{Riedi2003}  who worked out several important subcases; see also the article by S. Seuret \cite{Seur3}, who determined a criterium under which a function can be written as the composition of a time subordinator and a monoh\"older function, and   \cite{BaSe3} where J. Barral and S. Seuret  studied the multifractal spectrum of a L\'evy process, under a time subordinator  given by the  repartition function of a multifractal cascade.

\subsection{Multivariate  analysis of marathon physiological data} 

\label{MultivariateMar}

\begin{figure}
    \centering
    \includegraphics[scale=0.25]{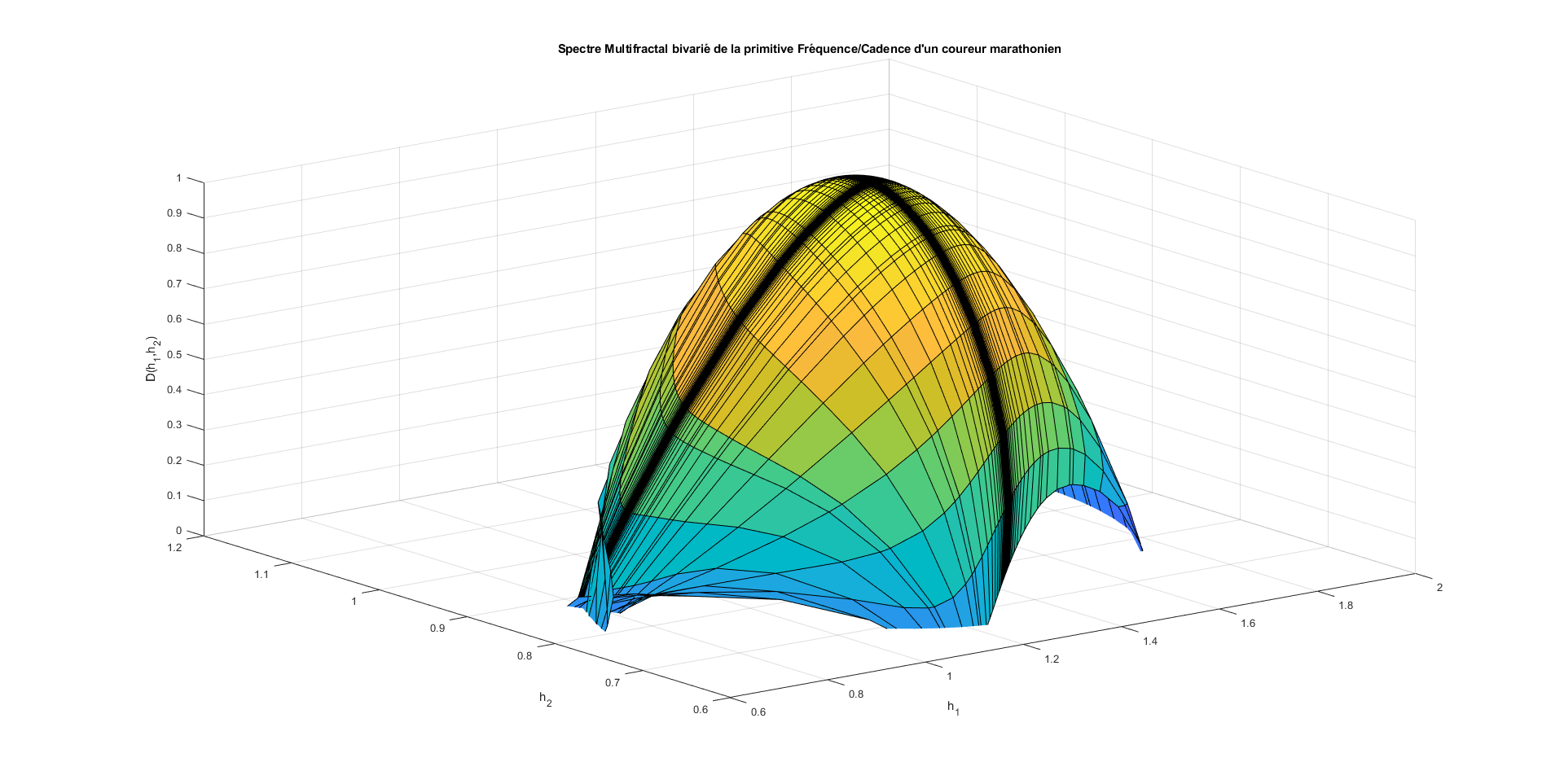}
    \caption{Representation of the bivariate H\"older Legendre spectrum of the primitives of heart beat frequency and cadence: this bivariate spectrum is derived from the same data that were used to derive   the two univariate spectra shown in Fig \ref{bilan_pl3}.  }
    \label{bilan_pl2}
\end{figure}

Let us consider  one of the marathon runners, and denote his heart beat frequency by  $f_f$  and his  cadence  by $f_c$ and by $f_f^{(-1)}$ and $f_c^{(-1)}$ their primitives. We performed the computation of the bivariate scaling function $\zeta_{f_f^{(-1)},f_c^{(-1)}}$ (using wavelet leaders)  and we show  its Legendre transform  ${ \cal L}_{f_f^{(-1)},f_c^{(-1)}}$ on Fig.   \ref{bilan_pl2}.
This spectrum  is widely spread, in strong contradistinction with the bivariate spectra obtained in  the previous section;  this indicates   that no clear correlations between the H\"older singularities of the primitives can be put in evidence. Fig. \ref{bilan_pl3} shows the two corresponding univariate spectra (which can be either computed directly, or obtained as projections of the bivariate spectrum).

\begin{figure}
    \centering
    \hbox{\hspace{-4em}\includegraphics[scale=0.35]{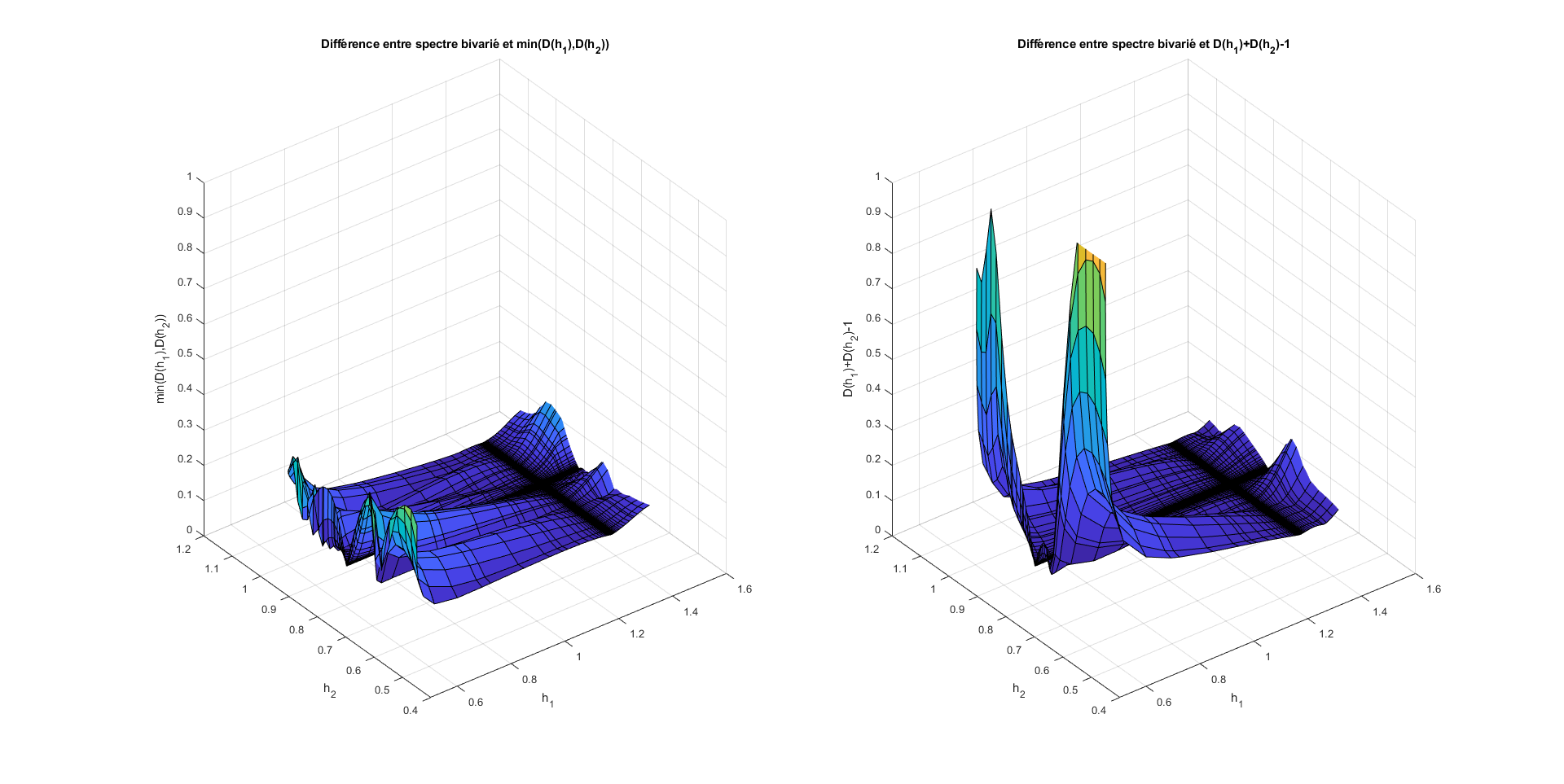}}
    \caption{Representation of the  difference of the bivariate spectrum and the two formulas proposed in \eqref{spectindep}  and \eqref{largint2}. The graph on the left is closer to zero, which suggests that the large intersection formula seems more appropriate in this case.}
    \label{bilan_biv1}
\end{figure}

 In order to test possible relationships  between the bivariate spectrum and the two corresponding  univariate  spectra, we compute the difference
\[ {\mathcal L} (H_1, H_2) - {\mathcal L} (H_1) - {\mathcal L} (H_2) +1, \]
which allows to test  the validity of  \eqref{spectindep} 
 and 
\[  { \cal L}_{(f_1,f_2)}  -   \min ( { \cal L}_{f_1} (H_1),  { \cal L}_{f_2} (H_2)),  \] 
which allows to test  the validity of  \eqref{largint2}, 
they are shown in Fig. \ref{bilan_biv1}.  This comparison suggests   that the large intersection formula is more appropriate  than the codimension formula in this case. 
Keeping in mind the conclusions of Section \ref{MultivariateSpec}, these results indicate that an hypothesis of both stationarity and independence for each signals is inapropriate   (indeed this would lead to the validity of the codimension formula), and on the opposite, these results are compatible with a pointwise regularity yielded by a limsup set procedure, as explained in Sec \ref{MultivariateSpec}. 

\section{Conclusion}  Let us give a summary of the conclusions that can be drawn from a bivariate multifractal analysis of data based on the Legendre transform method.  This analysis goes beyond the (now standard) technique  of estimating correlations of wavelet coefficients; indeed here wavelet coefficients are replaced by wavelet leaders, which leads to new scaling parameters  on which classification can be performed. On the mathematical side, even if the relationship between the Legendre and the multifractal spectra is not as clear as in the univariate case,  nonetheless, situations have been identified where this technique can either yield information on the nature of the singularities (e.g. the absence of oscillating singularities), or on the type of processes that can be used to model  the data (either of additive or of multiplicative type). In the particular case of marathon runners,  the present study shows a bivariate spectra between heart rate and cadence are related by the large intersection formula. In a recent study \cite{BillatPet} a multivariate analysis revealed that, for all  runners, RPE and respiratory frequency measured on the same runners during the marathon were close (their angle is acute on correlation circle of a principal component analysis) while the speed was closer to the cadence and to the Tidal respiratory volume at each inspiration and expiration). The sampling frequency of the respiratory parameters did not allow to apply the multifractal analysis which here reveals that the cadence and heart rate could be an additive process such as, possibly  a generalization of a Lévy process. Heart rate  and cadence are under the autonomic nervous system control and  Human beings optimize their cadence according his speed for minimizing his energy cost of running. Therefore, we can conclude that is not recommended to voluntarily change the cadence and this bivariate multifractal analysis mathematically shows that the cadence and  heart rate are not only correlated but we can conjecture that they can be modeled by  an additive process  until the end of the marathon. 

\label{conclu}

 \bibliographystyle{plain}
 \bibliography{IsaacRev.bib}
\end{document}